\documentclass{ar-1col}
\usepackage[numbers]{natbib}

\usepackage{amsfonts} 
\usepackage{amssymb} 
\usepackage{amsmath} 
\usepackage{verbatim} 
\usepackage[legalpaper,margin=1.5in]{geometry}

\jname{Xxxx. Xxx. Xxx. Xxx.}
\jvol{AA}
\jyear{YYYY}
\doi{10.1146/((please add article doi))}

\newcommand{\Kdf}[0]{\mathcal K_{\text{df},3}}

\newcommand{\df}[0]{\mathrm{df}}
\newcommand{\bKdf}[0]{\mathbf K_{\df}}
\newcommand{\bF}{{\mathbf F}}
\newcommand{\bG}{{\mathbf G}}

\newcommand{\bK}{{\mathbf K_2}}

\newcommand{\bKLth}{\mathbf K_{L,33}}

\newcommand{\bKdfth}{\mathbf K_{\mathrm{df}, 3 3}}

\newcommand{\bGK}{{ \mathbf G_{\mathbf K} }}
\newcommand{\bSig}{{\pmb \sigma^*}}
\newcommand{\bSigD}{{\pmb \sigma^{\dagger*}}}
\newcommand{\bA}{{\mathbf A}}

\newcommand{\bT}{{\mathbf T}}

\newcommand{\bZ}{{\mathbf Z}}

\newcommand{\Kiso}[0]{\mathcal K^{\rm iso}_{\text{df},3}}
\newcommand{\Kdfs}[0]{\mathcal K_{\text{df},3,s}}
\newcommand{\ieps}[0]{{i\epsilon}}
\newcommand{\NR}{{\rm NR}}
\newcommand{\on}{{\rm on}}
\newcommand{\PV}{{\rm PV}}
\newcommand{\cI}{\mathcal I}
\newcommand{\cK}{\mathcal K}
\newcommand{\cL}{\mathcal L}
\newcommand{\cM}{\mathcal M}
\newcommand{\cO}{\mathcal O}
\newcommand{\cS}{\mathcal S}

\begin{document}

\markboth{Hansen and Sharpe}{Lattice QCD and Three-particle Decays of Resonances}

\title{Lattice QCD and Three-particle Decays of Resonances}

\author{Maxwell T. Hansen,$^1$  and Stephen R. Sharpe$^2$
\affil{$^1$Theoretical Physics Department, CERN, 1211 Geneva 23, Switzerland; email: maxwell.hansen@cern.ch}
\affil{$^2$Physics Department, University of Washington, Seattle, WA 98195, USA; email: srsharpe@uw.edu}}

\begin{abstract}
Most strong-interaction resonances have decay channels involving three or more particles,
including many of the recently discovered $X$, $Y$ and $Z$ resonances.
In order to study such resonances from first principles using lattice QCD, 
one must understand finite-volume effects for three particles in the cubic box
used in calculations. Here we review efforts to develop a three-particle
quantization condition that relates finite-volume energies  to infinite-volume scattering
amplitudes. We describe in detail the three approaches that have been followed, 
and present new results on the relationship between the corresponding results.
We show examples of the numerical implementation of all three approaches and
point out the important issues that remain to be resolved.
\end{abstract}

\maketitle

\tableofcontents

\section{INTRODUCTION}

One of the striking features of the strong interaction is the abundance of resonances.
These are very short-lived (having widths $\Gamma\sim 100\;$MeV), and
are not asymptotic states, but are manifested through the behavior of the scattering
amplitudes of particles that are stable under the strong interactions, e.g. pions, kaons
and nucleons.
Examples of resonances include the rho, decaying via $\rho \to \pi \pi$, 
as well as the $a_1(1260) \to \pi \pi \pi$ and the Roper resonance, $N(1440) \to N \pi, N \pi \pi$, 
where in each case we have shown the final states with the highest branching fractions. 

The lightest such resonances form patterns when arranged according to basic properties
such as spin, charge and strangeness, for example the flavor SU(3)
decuplet of $J=3/2$ baryons ($\Delta$, $\Sigma^*$, $\Xi^*$, and $\Omega$) and  
the nonet of $J^P=1^-$ mesons ($\rho$, $\omega$, $K^*$, and $\phi$).
This regularity, together with that of the particles stable under strong decay,
was crucial in determining the underlying degrees of freedom, the quarks and gluons.
This culminated in the formulation of quantum chromodynamics (QCD) in the early 
1970s~\cite{GrossWilczek,Politzer,FGML}.

While the lightest resonances can be categorized in quark-model language
as quark-antiquark and three-quark states, this does not hold for higher-lying
states. Examples include the $f_0(980)$ and $a_0(980)$, long considered
to have a tetraquark ($q q \bar q\bar q$) component~\cite{Jaffe},
and pentaquark candidates such as the $P_c(4450)^+$~\cite{pentaquark}.
Other resonances that do not fit into the simple quark-model classification
are the recently discovered $X$, $Y$ and $Z$ states, which contain open
or hidden charm and bottom quarks.
For a recent summary of the experimental and theoretical
status of the many states whose classification is unclear, 
see Ref.~\cite{Karliner} and the Particle Data Group listings and reviews~\cite{PDG18}.

This situation, and in particular the multitude of $X$, $Y$ and $Z$ resonances,
has led to a resurgence of interest in hadron spectroscopy,
and a renewed appreciation of the need to extract predictions from first-principles QCD. 
In particular, many of the new states lie close to thresholds, and involve decays to
multiple channels, and it is crucial to disentangle kinematical effects such as threshold
cusps from truly resonant behavior.
The latter is identified as a pole in the analytic continuation of a scattering amplitude to complex values of the center-of-mass energy.

A crucial tool in disentangling the underlying properties of such resonances
is first-principles calculations based in lattice QCD (LQCD).
With this method, unlike with the quark model or other approximate approaches, 
one can systematically remove all sources of uncertainty in the calculations.
The major such sources are the statistical errors inherent in
a Monte Carlo calculation, the need to work at nonzero lattice spacing,
the use of larger-than-physical quark masses,
and the need to work with a finite space-time volume, with spatial box length\footnote{%
Most LQCD calculations use cubic spatial boxes, and we consider only this case in
this review.}
 $L$  and Euclidean time extent $L_t$.
Over the last decade or so, an increasing number of LQCD results for single-particle
quantities have controlled all of these errors, in some cases at subpercent precision.
One indication of this improved level of control
is the increasingly common use of physical light-quark masses in calculations.
For a review of results for well-controlled single-particle quantities using LQCD, 
see Ref.~\cite{FLAG3}.
\begin{marginnote}[]
\entry{Lattice QCD}{Regularization suited to systematic numerical calculations.}
\end{marginnote}

Using LQCD to calculate the properties of resonances is, however, more challenging
than for single-particle properties.
A resonance is observed experimentally by studying the scattering of the decay products, e.g.~two pions in the case of the $\rho$ resonance.
By measuring scattering rates for various kinematics and then performing a partial-wave analysis, one can in principle determine the scattering amplitudes projected to any given angular-momentum component and search for resonances. 
Lattice calculations cannot, however, reproduce this setup.
The use of a finite volume does not allow the consideration of states with
well-separated decay products (so one cannot approach the in- and out-states needed
for a theoretical description of scattering in quantum field theory)
and the need to use Euclidean time (in order to avoid a numerically intractable
sign problem) makes real-time processes such as scattering inaccessible.
Thus one is forced to use an indirect approach.

The indirect approach that is by now widely applied was first introduced in seminal work by 
L\"uscher~\cite{Luscher1,Luscher2,Luscher3}.
The essential observation is that the volume dependence of the energies of multiparticle
states is governed by infinite-volume scattering amplitudes. By determining,
in a lattice calculation, the finite-volume spectrum as a function of $L$, one is thus doing
something analogous to a scattering experiment. Crudely speaking, the multi-particle finite-volume
state in a large enough box contains particles that are almost moving freely,
and thus approximate a scattering state.

In the case of resonances that only decay to channels containing two particles,
this has been placed on a rigorous footing by the derivation of so-called 
two-particle quantization conditions, i.e.~equations that are satisfied only
at the energies of finite-volume states and yet depend on
infinite-volume scattering quantities. The simplest cases were worked out in
Refs.~\cite{Luscher1,Luscher2,Luscher3} (and will be reviewed
in subsequent sections), and extensions to arbitrary spins, 
noncubic boxes, moving frames,
and multiple two-particle channels have been subsequently derived. We do not
review this literature here, as it is not the topic of this work, but point the interested
reader to the comprehensive review provided in Ref.~\cite{Raulreview}.
\begin{marginnote}[]
\entry{L\"uscher's method:}{Determine resonance properties from finite-volume spectrum}
\end{marginnote}

We now come to the essential phenomenological motivation for the present review:~Most resonances have some decay channels that involve three or more stable hadrons.
Examples noted above are the $a_1(1260)$, with a dominant decay into three pions,
 the Roper resonance, which decays both to two- and three-particle channels,
and many of the $X$, $Y$ and $Z$ resonances.
For such resonances, the two-particle formalism simply does not apply.
Thus, in order to address many pressing questions in hadron spectroscopy using LQCD,
a three-particle quantization condition is needed. 
This is an equation that, given information about two- and three-particle scattering, 
predicts the finite-volume spectrum
or, conversely, provides constraints on the scattering amplitudes given the
spectrum calculated using LQCD.\footnote{%
As we will see in subsequent sections, the connection between finite-volume spectrum
and scattering amplitudes is more indirect in the three-particle case compared to that
for two particles.}
The major purpose of this review is to explain the theoretical progress
that has been made over the last five or so years in deriving
three-particle quantization conditions.

Another motivation for this review is that advances in algorithms, methods,
and the speed of computers,
have allowed LQCD calculations to determine the finite-volume spectrum
in the energy regime in which states have a significant three-particle component.
Many examples can be found in Ref.~\cite{Raulreview},
but we note in particular the spectra determined in Refs.~\cite{HADSPEC} 
and~\cite{BulavaMorningstar}. 
At present, these calculations use heavier-than-physical quarks 
(so that $m_\pi \gtrsim 230\;$MeV),
but, nevertheless, their interpretation requires a quantization condition
that can account for three, and in some cases more, particles.
Indeed, the need for such a quantization condition becomes more pressing
as the quark masses are lowered to their physical values, for then the
multi-pion thresholds drop rapidly.

\bigskip
Although motivated by results from LQCD, the derivations of the
quantization conditions are in fact based in finite-volume continuum quantum field theory (QFT).
In particular, we assume in the following that calculations have controlled the
errors associated with nonzero lattice spacing and unphysical quark masses by
appropriate extrapolations (or, in the latter case, possibly interpolations). 
As noted above, there are also errors associated with the finite extent of the lattice
in the Euclidean time direction, $L_t$.
Lattice calculations are mostly done with boundary conditions in the temporal direction chosen
to give the system a thermal interpretation, with temperature $T=1/L_t$.
By choosing $L_t$ large enough one works at very low temperatures,
and then it is possible to extract the spectrum with controlled errors from the dependence of 
correlators on the Euclidean time separation. 
Thus the only issue we address here
is the impact of working in a finite, cubic box, of length $L$. 
We further assume that, as in most lattice calculations, the spatial boundary conditions are periodic.
\begin{marginnote}[]
\entry{Box length, $L$}{Periodicity of the finite-volume in each of the three spatial directions}
\end{marginnote}

Within this box one can fix all internal quantum numbers of the overall state.
For example, one can set $Q= |e|$, $B=S=0$ ($B$ being baryon number, 
and $S$ strangeness), choose odd G-parity,\footnote{%
To keep the examples discussed in this paragraph simple,
we work in the limit of exact isospin symmetry.}
and also fix the total
three-momentum $\vec P$ to one of the allowed finite-volume values $2\pi \vec n/L$ 
(with $\vec n$ a vector of integers).
Then the lightest finite-volume state corresponds to a $\pi^+$ with momentum $\vec P$, and the
excitations correspond to interacting
$\pi^+ \pi^+ \pi^-$ and $\pi^+\pi^0\pi^0$ states.
If instead one projects onto $Q=2 |e|$ and even G-parity, while keeping $B=S=0$, then the
available states are approximately described as interacting 
$\pi^+\pi^+$, $\pi^+ \pi^+\pi^+\pi^-$ and  $\pi^+\pi^+\pi^0\pi^0$ states, etc. 
Similarly, if we take $Q=3|e|$ and odd G-parity, 
then the lightest state consists of $\pi^+\pi^+\pi^+$.

It is important to keep in mind, however, that we
do not need to keep track of the individual particle components.
We simply create the state with an operator having the requisite quantum numbers. Then
the QCD dynamics, encoded in terms of quarks and gluons interacting via the QCD Lagrangian,
lead to a low-lying spectrum of multi-hadron finite-volume states. 
The only output is a set of $L$-dependent energy levels, and this is all the input needed by the
quantization conditions.

It will be useful for the more technical discussion of the subsequent sections to
describe some general features of the dependence of energy levels on $L$.
For single, stable particles, a key result is that the energies, and other properties,
depend on volume as $(M_\pi L)^{-n} \exp(- \alpha M_\pi L) $, 
where $n$ and $\alpha$ depend on the quantity~\cite{Luscher1}.
These corrections arise from virtual pions propagating to the neighboring cells of the periodic system. They drop off rapidly with $L$, and it has been found that using $M_\pi L \gtrsim 4$ is usually sufficient for finite-volume uncertainties to become subleading to other sources.
Throughout this review, we assume that such exponentially-suppressed dependence
can be neglected. The weakness of such dependence is crucial for the above-described
successes of LQCD in attaining subpercent precision for certain single-particle quantities.
\begin{marginnote}[]
\entry{Key approximation:}{Neglect of exponentially-suppressed volume dependence}
\end{marginnote}

The volume dependence for multiple-particle states is, however, quite different. In this case the asymptotic behavior exhibits power-law scaling of the form $L^{-n}$. 
We describe the origin of this behavior below, but for now note only that multiparticle finite-$L$ effects fall off much more slowly than the exponentially suppressed terms and cannot be ignored. Indeed, the quantization conditions allow this dependence to be used as a tool
rather than an unwanted artifact. 
In this way, an apparent source of systematic uncertainty in LQCD calculations has become a powerful window into resonance physics.

\bigskip

This review is organized as follows. Three approaches have been followed in the
development of the three-particle quantization condition, 
and we discuss them in turn.\footnote{%
We also note the pioneering work of Ref.~\cite{AkakiPolejaeva},
which showed that the three-particle
spectrum depends only on S-matrix elements.
Another early step towards a quantization condition was taken in Ref.~\cite{RaulZohreh}.}
The first uses a diagrammatic, all-orders analysis in a generic effective QFT.
We refer to this as the RFT approach, 
with the R emphasizing that this is a relativistic approach.
It was developed by us in Refs.~\cite{HS1,HS2} for the simplest case of three identical
scalars with a $Z_2$ symmetry forbidding $2\leftrightarrow 3$ transitions, and
subsequently generalized in collaboration with Brice\~no in Refs.~\cite{BHS2to3,BHSK2poles}.
This is the topic of the following section, Sec.~\ref{sec:RFT}.
We first describe the derivation of the two-particle quantization condition in Sec.~\ref{sec:QC2},
providing two simple examples in order to motivate the general derivation.
We then, in Sec.~\ref{sec:QC3}, provide a description of the derivation of the
three-particle quantization condition. Here we are able to use results from
Ref.~\cite{BHSK2poles} to simplify and shorten the derivation given in the original
work, Ref.~\cite{HS1}. Our hope is that this will make this rather technical derivation
more accessible. 
We close Sec.~\ref{sec:RFT} with brief discussions of how the three-particle quantization
condition can be truncated and thus made practical, and of two analytic checks
of the formalism.

Two alternate approaches have subsequently been followed, both of which
greatly simplify the derivation of the quantization condition. These are described
in Sec.~\ref{sec:alt}. To date, these only consider the case in which the interaction
between particle pairs (denoted ``dimers'') is purely $s$-wave. This is in contrast
to the RFT approach, for which all waves are included in the dimer interactions.
The two alternate approaches are that based
 on non-relativistic effective field theory (NREFT)~\cite{Akaki1,Akaki2},
 and that using a finite-volume implementation of unitarity constraints,
 which we refer to as the finite-volume unitarity (FVU) approach~\cite{MD1,MD2}.
 For both approaches we describe the derivation of first the two- and then the three-particle
quantization conditions, and then describe the relation of the results to those in the RFT
approach. These latter two sections, Secs.~\ref{sec:NREFTtoRFT}
and \ref{sec:FVUtoRFT}, present new results, which we think illuminate the
similarities and differences between the approaches.

A key question for all approaches is whether they can be implemented in practice.
This has been addressed over the last year or so in all three approaches by showing
how, in the simplest approximation of $s$-wave dimer interactions and a local three-particle
interaction, the quantization conditions can be used to predict the finite-volume
spectrum. Sec.~\ref{sec:num} gives examples of the results from all three approaches.

We close with a list of summary points and issues for future work.

\bigskip

In order to keep this review within bounds, there are many topics that we do not cover. 
As already noted, we do not discuss, except in passing, the two-particle formalism,
or the extensive numerical results from LQCD calculations using the two-particle quantization condition to determine
two-particle scattering amplitudes.
For a recent, thorough review of both of these topics, we refer the reader again to
Ref.~\cite{Raulreview}.

We also do not discuss the approach to two- and three-particle systems developed by
the HALQCD collaboration, and reviewed in Ref.~\cite{HALreview}.
In the two-particle case,
this approach uses the Bethe-Salpeter amplitude in order to extract a potential that can
then be inserted into the  Schr\"odinger equation to determine bound-state energies
and scattering amplitudes. It has been successfully applied to many two-baryon systems---for
a recent example see Ref.~\cite{HALrecent}.
The extension to three particles~\cite{HAL3} is, 
however, so far restricted to the nonrelativistic domain,
and thus is not directly applicable to most of the resonances of interest in QCD.

We also do not discuss the impressive recent progress in simulating 
multiple interacting nucleons by discretizing the truncated pionless
chiral EFT, and working in a finite volume. For a review see Ref.~\cite{Ulfreview},
and for an example of recent work see Ref.~\cite{Ulfrecent}.
This is a powerful method for studying bound states and near-threshold
behavior, but does not apply to the resonances of interest, where dynamical pions
are essential.

Finally, we do not discuss recent  ideas for determining the finite-volume
multi-particle spectrum in NRQM using a variational approach~\cite{variational},
as again this is restricted to the nonrelativistic regime.

\section{RELATIVISTIC QUANTIZATION CONDITIONS USING A DIAGRAMMATIC APPROACH IN QFT}
\label{sec:RFT}

In this section we review the derivations 
of the two- and three-particle quantization conditions 
 using an all-orders diagrammatic analysis in a generic relativistic field theory.
We refer to this as the RFT approach.
We begin with two particles as this allows us to explain the general strategy
and to introduce notation that will aid in the explication of
the more complicated case of three particles.
We will derive these results in the simplest setting, namely for identical, spinless
particles, which is the only setting considered to date for the three-particle
quantization condition.

\subsection{Two-particle quantization condition in the RFT approach}
\label{sec:QC2}

The two-particle quantization condition provides the relation between the spectrum
of a field theory in a finite box and the infinite-volume two-to-two scattering
amplitude, $\cM_2$. We begin with some kinematical notation for $\cM_2$.
We use $P^\mu=(E,\vec P)$ 
to denote the total 4-momentum of the scattering pair, so that $P^\mu P_\mu = E^2 - \vec P^2 = s$.
Denoting the 4-momentum of one of the incoming particles as $k$, and that of one of the outgoing as $k'$,
we write the dependence of the amplitude as $\cM_2(P;k', k)$.
We use this notation also for off-shell amplitudes, for which $k^\mu k_\mu \ne m^2$, with $m$ the physical mass of the particle.\footnote{%
These are defined as the sum of all two-to-two amputated diagrams using the fundamental
field of the theory as the external operator. Field redefinitions can change these
amplitudes off shell, but the on-shell values, which are all that enter our final expressions,
are invariant.}
A special role is played by the c.m.~(center-of-momentum) frame. 
We denote quantities in this frame with a superscript $*$, e.g.
the total energy is $E^*=\sqrt{s}$, and the incoming 3-momentum $\vec k$ boosted
to this frame is $\vec k^*$. When the two incoming particles, with momenta $k$ and $P-k$, are each on shell, this implies a constraint on the magnitude of the CM-frame momentum, $k^*\equiv |\vec k^*|$. In this case one finds
that the latter is equal to $q^*=\sqrt{E^{*2}/4 - m^2}$, meaning that for fixed $E^*$ only angular degrees of freedom remain.
This allows one to decompose the on-shell amplitude 
into angular-momentum components,
$\cM_2^{(\ell)}(s)$, in the standard way.
\begin{marginnote}[]
\entry{Superscript *}{Denotes quantities defined in the c.m.~frame.}
\end{marginnote}

Unitarity provides an important constraint on $\cM_2^{(\ell)}(s)$, one that will play
a central role in Sec.~\ref{sec:FVU2}. Specifically, it implies that, for $s\ge 4 m^2$,
\begin{equation}
\mathcal M_2^{(\ell)}(s)^{-1} = \mathcal K_2^{(\ell)}(s)^{-1} - i \rho(s) \,,\quad
\mathcal K_2^{(\ell)}(s)^{-1} = 
\frac{q^* \cot \delta^{(\ell)}(q^*)}{16 \pi \sqrt{s}} \,,\quad
\rho(s) = \frac{q^*}{16\pi\sqrt{s}}\,,
\label{eq:unitarity}
\end{equation}
\begin{marginnote}[]
\entry{$\mathcal K_2^{(\ell)}(s)$}{Relation between K matrix and scattering amplitude}
\end{marginnote}
where the K matrix,  $\mathcal K_2^{(\ell)}(s)$, is a real function that is meromorphic (analytic up to poles) for $s > 0$. 
When we discuss the three-particle case we
need to consider $\cM_2^{(\ell)}(s)$ below threshold such that $q^{*2} < 0$. To make sense of this is useful to note that $\cM_2^{(\ell)}(s)$ has a branch cut along the real axis in the complex $s$ plane, running from $s = 4 m^2$ to $\infty$. The conventions established above correspond to real energies just above the cut. To remain on the same Riemann sheet for $s<4m^2$, one must analytically continue the phase-space factor as $\rho(s) = i |q^*|/(16\pi\sqrt{s})$. 

Turning now to the finite volume, we open with only a brief comment on our set up: In this review we restrict attention to periodic, cubic volumes, with length $L$
in each of the three spatial directions.

\subsubsection{Example: leading term in the threshold expansion\label{sec:ex1}}
To gain intuition into the general approach, we discuss two simple examples of 
the derivation of the quantization condition. The first concerns the lowest 
two-particle energy for $\vec P=0$, which, as was already shown over
60 years ago~\cite{HY}, satisfies 
\begin{equation}
E_0(L) = 2 m + {4 \pi a}/{ (m L^3)} + \mathcal O(1/L^4) \,.
\label{eq:HYth}
\end{equation}
\begin{marginnote}[]
\entry{Threshold expansion}{Leading order term}
\end{marginnote}
Here $a$ is the scattering length, defined by
\begin{equation}
q^* \cot \delta^{0}(q^*) = - {1}/{a} + \mathcal O(q^{*2}) \,.
\label{eq:a}
\end{equation}

\begin{figure}
\begin{center}
\includegraphics[width=\textwidth]{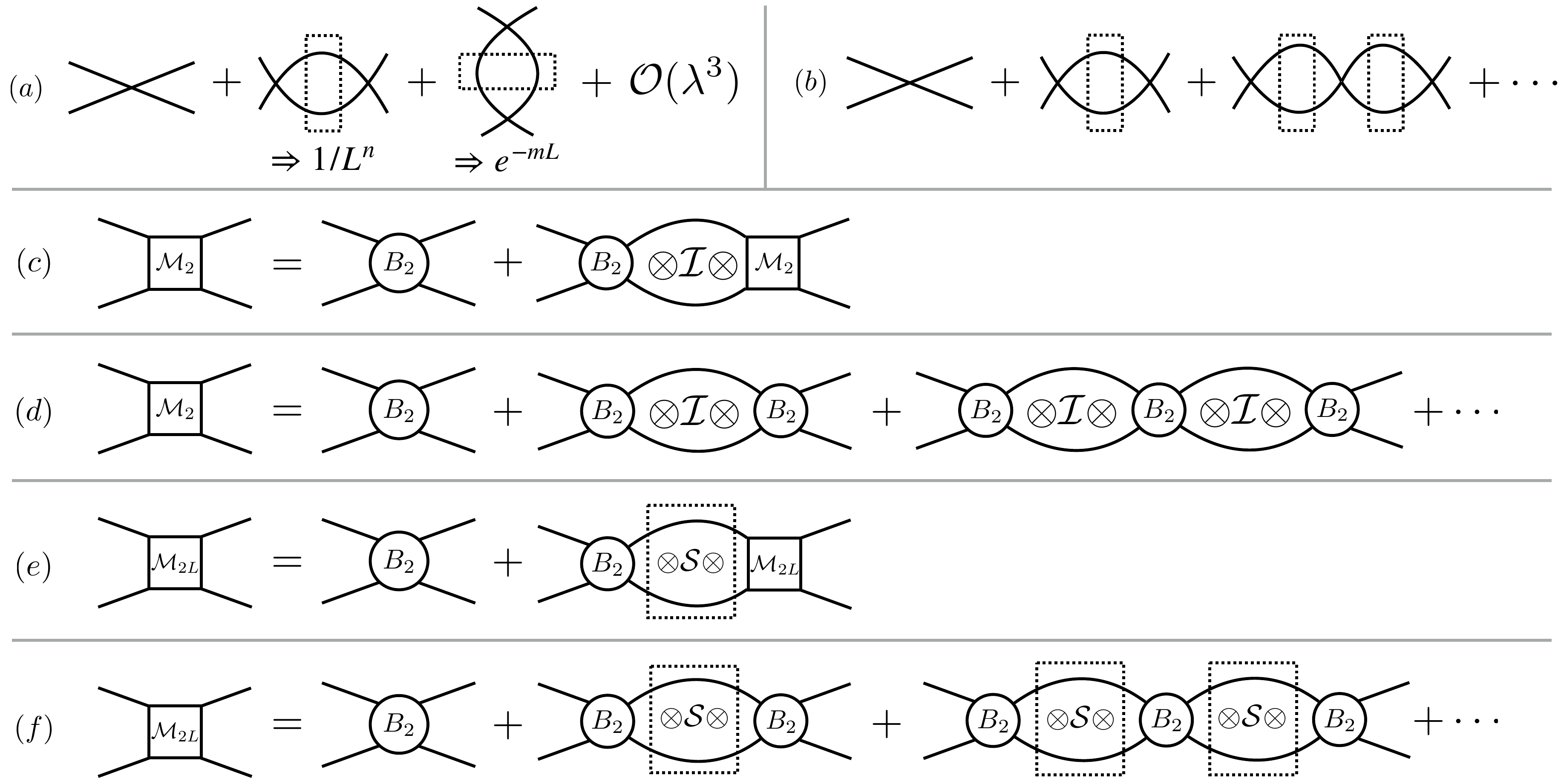}
\caption{Representations of the scattering amplitude, $\cM_2$ and its finite-volume
correspondent, $\cM_{2L}$. (a) First- and second-order contributions to $\cM_{2L}$
in $\lambda\phi^4$ theory, with the dashed rectangles indicating that the spatial
momenta are summed over the discrete finite-volume values.
(b) The set of diagrams that must be summed in order to calculate the leading
order energy shift in $\lambda\phi^4$ theory.
(c) Bethe-Salpeter (BS) equation for the infinite-volume amplitude, with $B_2$ being the
BS kernel. (d) Iterated version of the BS equation. (e) BS equation for the finite-volume
amplitude, with $\cS$ and the dashed rectangle both indicating a sum over finite-volume
three-momenta. (f) Iterated version of finite-volume BS equation.}
\label{fig:1plus2}
\end{center}
\end{figure}

It is convenient here, and in the following,
to introduce a finite-volume correlator that is closely related to the two-to-two scattering amplitude, $\cM_2$.  This object, called $\cM_{2L}$, will have poles at the energies of
the finite-volume states. It is defined by calculating
exactly the same set of Feynman diagrams as for $\cM_2$,
but with a sum rather than an integral over the allowed
 three-dimensional momentum modes. 
 For example, in $\lambda \phi^4$ theory, the leading-order and next-to-leading-order contributions are shown in {\bf Figure~\ref{fig:1plus2}(a)} and given by
\begin{equation}
i \mathcal M_{2L} \equiv - i \lambda - \frac{\lambda^2}{2} \int \frac{d \ell^0}{2 \pi}  \frac{1}{L^3} \sum_{\vec \ell} \frac{i}{\ell^2 - m^2 + i \epsilon} \frac{i}{(P - \ell)^2 - m^2 + i \epsilon} + \cdots + \mathcal O(\lambda^3) \,,
\label{eq:M2Lth}
\end{equation}
where the ellipses represents the $t$- and $u$-channel diagrams. 
 Some choice of UV regularization is implicit here, but need not be specified
as it will play no role in the final result.

Our aim is now to pull out the power-law volume dependence.
If all sums were replaced by integrals, then we would simply obtain 
the second-order expression for $\cM_2$.
For the $t$- and $u$- channel contributions, it can be shown, as discussed below,
that the sum-integral difference has an
exponentially-suppressed volume dependence, scaling as $e^{- m L}$.
We neglect such dependence throughout this review as it is numerically
small in practice.
Thus the only source of power-law volume dependence is the $s$-channel loop
shown explicitly in Eq.~(\ref{eq:M2Lth}).
This can be simplified by doing the $\ell^0$ integral, and then replacing the
sum with the integral plus sum-integral difference. 
One then obtains
\begin{equation}
i \mathcal M_{2L} =  i \mathcal M_2 - \frac{\lambda^2}{2}  \bigg [ \frac{1}{L^3} \sum_{\vec \ell} - \int \! \frac{d^3 \vec \ell}{(2 \pi)^3} \bigg ] \frac{i}{2 \omega_{  \ell} 2 \omega_{  P -   \ell} (E -  \omega_{  \ell} - \omega_{  P -   \ell} + i \epsilon)}  + \mathcal O(e^{- m L},\lambda^3) \,,
\label{eq:M2LM2}
\end{equation}
where $\omega_{  \ell} \equiv \sqrt{m^2 + \vec \ell^2}$.

We next set $\vec P=0$ and consider the weak coupling limit, $\vert \mathcal M_2 \vert \ll 1$.
Then the two-particle energy will be very close to the lowest-lying non-interacting state,
$E=2m + \cO(\lambda)$.
In this regime,
the dominant volume-dependent contribution to $\cM_{2L}$
comes from the $\vec \ell=0$ contribution to the sum, leading to
\begin{equation}
\label{eq:M2Lnearpole}
i \mathcal M_{2L} =  i \mathcal M_2 - \frac{\lambda^2}{2}    \frac{1}{L^3}   \frac{i [1 + \mathcal O(1/L)]}{4 m^2 (E -  2 m + i \epsilon)}  + \dots  
\end{equation}
At this stage we identify a problem with the truncated expression. While the leading-order term contains no poles, the next-to-leading order result contains a pole at the non-interacting level $E = 2m$. Both results are incorrect for the interacting theory. 
The problem arises because, 
for $E - 2 m = \mathcal O(\lambda)$, the first two terms are effectively of the same order.
More importantly, some of the terms we have neglected are of this order as well.

To resolve the issue we must include all diagrams with any number of $s$-channel bubbles,
as shown in {\bf Figure~\ref{fig:1plus2}(b)}
Doing so leads to a geometric series, and summing this leads to a shift in the pole in $\mathcal M_{2L}$.
To see this explicitly, we work to all orders in $\lambda/(E - 2 m)$ while only working to leading-order in the scattering amplitude
(so that $\cM_2=-\lambda$). We find 
\begin{equation}
\label{eq:M2LnearpoleSUM}
 \mathcal M_{2L} =     \mathcal M_2   \sum_{n=0}^\infty
  \bigg (      \frac{(-1)}{2 L^3}   \frac{[1 + \mathcal O(1/L)]}
  {4 m^2 (E -  2 m + i \epsilon)} \mathcal M_2  \bigg )^n 
  = \frac{\cM_2 (E - 2 m)}{ E - 2 m     +\cM_2/(8m^2 L^3) + \mathcal O(1/L^4) } \,.
\end{equation}
\begin{marginnote}[]
\entry{Threshold shift}{Example of a shifted pole in the finite-volume correlator}
\end{marginnote}
If we now make the replacement  $\mathcal M_2 = -32 \pi m a$,
valid at leading order in $\lambda$ for all $s$ (and to all orders in $\lambda$ for $s = 4 m^2$), then we find that the pole is indeed shifted
to the position given in Eq.~(\ref{eq:HYth}).

\subsubsection{Example: quantization condition in $1+1$ dimensions\label{sec:1plus1}}
Our second example shows how the quantization condition emerges for general
values of $E$ and $L$.  Working in $1+1$ dimensions keeps the essential features
of the derivation while simplifying the calculation.
For simplicity, we restrict to the c.m.~frame,  $\vec P=0$, and drop the argument $P$
in $\cM_2$ and related quantities.

It is instructive to first show how the unitarity relation arises diagrammatically.
To this end, we expand $\cM_2$ in powers of the Bethe-Salpeter (BS) kernel,
as shown in {\bf Figures~\ref{fig:1plus2}(c)-(d)}. This kernel is the sum of all diagrams that 
are two-particle irreducible in the $s$-channel. 
Thus it contains loops with three or more particles, and single-particle poles, but no two-particle loops. We focus on the kinematic region $m < E < 3 m$,
within which the BS kernel has no on-shell cuts, and thus is real.
Thus the only sources of imaginary contributions are the two-particle loops,
which can have on-shell cuts.
Performing the time component integrals, the expression for $\cM_2$ can be brought
into the form\footnote{%
This requires a redefinition of $B_2$ to absorb the ``Z-diagram'' contribution 
from two-particle loop, as well as the residue function of the remaining pole.}
\begin{align}
\hspace{-10pt} i \mathcal M_2(p',p)  & =  \sum_{n=0}^\infty \prod_{j=0}^{n-1} \left (  \frac12 \int_{k_{j+1}}  i B_2(k_j , k_{j+1}) \frac{i}{(2 \omega_{k_j})^2 (E - 2 \omega_{k_j} + i \epsilon)}   \right ) i B_2(k_n, p) \bigg \vert_{k^0=p'} \,.
\label{eq:M21dim}
\end{align}
Here we have left the momenta $p'$ and $p$ general. We now set these on-shell, via $p'=p=q^*$, and drop the dependence on the left-hand side.

We are interested in extracting the imaginary part. Thus we use the
identity relating the $i\epsilon$ prescription to a principal value (PV) prescription plus
an imaginary $\delta$-function term:
\begin{align}
\cM_2  & =  \sum_{n=0}^\infty  
\left (   -  B_2 \otimes \left[\mathcal I_{\PV} -  \frac12 
\int_k  \frac{i}{(2 \omega_{k})^2} \delta(E - 2 \omega_{k})\right]  \otimes \right )^n  B_2 \,,
\end{align}
where $\otimes$ indicates integration of adjacent quantities over the common momentum, 
and $\cI_\PV$ is the pole term with the PV prescription.
Thus factors of $B_2$ adjacent to $\otimes$ are evaluated off shell.
By contrast, those adjacent to the delta-function can be set on shell, 
i.e.~with $2  \omega_{k_j} = E$. 

Performing the integral we find
\begin{align}
\cM_2  & =   \sum_{n=0}^\infty  \left ( [  -  B_2 \otimes \mathcal I_{\PV} \otimes ] \  
+  B_2   \,   \frac{i  }{8 p E}  \right )^n  B_2 \,.
\end{align}
The sum on the right-hand side with the imaginary term dropped leads precisely
to the K matrix, $\cK_2$, where we stress that the external arguments of the $B_2$s
are on shell. Reorganizing the full right-hand side, keeping the imaginary
term, thus gives
\begin{align}
\cM_2  & =  
\sum_{n=0}^\infty  \left ( \sum_{m=0}^\infty [  -  B_2 \otimes \mathcal I_{\PV} \otimes ]^m B_2
\frac{i}{8q^*E}  \right )^n  \sum_{p=0}^\infty [  -  B_2 \otimes \mathcal I_{\PV} \otimes ]^p B_2\,,
\\
&  =    \sum_{n=0}^\infty \left (  - \mathcal K_2   \left[ - \frac{i  }{8 q^* E} \right ]   \right )^n   \mathcal K_2  
 = 
 \frac1{\cK_2^{-1} - i/(8 q^* E)}\,,
\end{align}
\begin{marginnote}[]
\entry{Unitarity of $\cM_2$}{$1+1$-dimensional case}
\end{marginnote}
where again it is crucial that the $B_2$s next to the $i/(8q^* E)$ terms are on shell,
so that it is the on-shell $\cK_2$ that appears.
The final result is simply the unitarity relation, 
Eq.~(\ref{eq:unitarity}), but written in one spatial dimension.

We now argue that a similar analysis can be applied to the finite-volume correlator $\cM_{2L}$
introduced above. We recall that $\mathcal M_{2L}$ 
is a real function of energy whose poles give the finite-volume spectrum  of the theory. 
The first step is to note that  $\cM_{2L}$ is given by Eq.~(\ref{eq:M21dim}),
except that the integrals are replaced by finite-volume sums over $k_j = 2\pi n_j/L$,
with $n_j$ an arbitrary integer, and the $i\epsilon$ is dropped.
This is shown diagrammatically in {\bf Figure~\ref{fig:1plus2}(e)-(f)}.
The BS kernels in $\cM_{2L}$ are, strictly speaking, the finite-volume versions, but
these differ from those in infinite volume only by exponentially-suppressed terms,
a difference we neglect.
This holds in our kinematic regime because the loops inside $B_2$ have integrands
that cannot go on shell, and are thus nonsingular. One can then use the Poisson
summation formula to show that the sum-integral difference is exponentially suppressed.
This result holds in any number of dimensions, and was first derived
in Ref.~\cite{Luscher2}. 
\begin{marginnote}[]
\entry{Key result}{BS kernels have exponentially-suppressed $L$ dependence}
\end{marginnote}

We now follow analogous steps to the demonstration of unitarity shown above, 
except that here we separate the sum over finite-volume modes 
into the principal-value integral and the residual sum-integral difference
(instead of separating the $i \epsilon$-integral into a PV integral and an imaginary part):

\begin{align}
\mathcal M_{2L} 
&  = \sum_{n=0}^\infty \left ([  -  B_2 \otimes \mathcal I_{\PV} \otimes ] \  
-  B_2 \otimes  \, \frac12 \bigg [ \frac{1}{L} \sum_k - \PV\int \frac{dk}{2 \pi} \bigg ]
\frac{ 1}{(2 \omega_k)^2 (E-2\omega_k)}  \otimes  \right )^n B_2 \,, \\
&  =  \sum_{n=0}^\infty \left ([  -  B_2 \otimes \mathcal I_{\PV} \otimes ] \  -  B_2  \,   \frac{ 1}{4E L} \frac{L^2}{4 \pi^2} \bigg [   \sum_n -  \PV\int  dn \bigg ]\frac{1  }{  x^2 - n^2 }    \right )^n B_2 + \mathcal O(e^{- m L})  \,,
\end{align}
where in the second line we have rearranged the expression and introduced 
$x \equiv q^* L/(2 \pi)$. 

Unlike in our derivation of the unitarity relation, here we have no Dirac delta function to project $B_2$ to its on-shell value. Nonetheless, we note that the factors of the BS kernels in the $L$-dependent terms can, in fact be evaluated at $k = q^*$. This is justified because the difference is exponentially suppressed 
\begin{multline}
  \bigg [ \frac{1}{L} \sum_k - \PV \int \frac{dk}{2 \pi} \bigg ]\frac{  \omega_k}{(2 \omega_k)^2 (q^{*2} - k^2)}     [ B_2(p'', k)  B_2(k, p')  -  B_2(p'', q^*)  B_2(q^*, p') ]  = \\
  \bigg [ \frac{1}{L} \sum_k -  \int \frac{dk}{2 \pi} \bigg ]\frac{  \omega_k}{(2 \omega_k)^2 (q^{*2} - k^2)} \big [  b(k)   (q^{*2} - k^2) \big ] \propto
   \sum_{n'\ne 0}  \int \frac{dk}{2 \pi} \frac{b(k) e^{i L k n'}}{\sqrt{k^2 + m^2}} 
   = \mathcal O(e^{- m L}) \,.
\end{multline}
\begin{marginnote}[]
\entry{On-shell projection}{from sum-integral difference}
\end{marginnote}
The first line here shows the difference between on- and off-shell BS kernels 
and, on the second line, we have used the result that this must equal a smooth function 
(denoted $b(k)$) times $q^{*2} - k^2$. 
Canceling the pole, and then using the Poisson summation formula, we have rewritten the result as a series of Fourier transforms with respect to integer multiples of $L$. This leads to our conclusion that kernels can be placed on shell up to terms that we neglect. In this way we have an effective delta function, in place of the true factor of $\delta(E - 2 \omega)$ that appeared in the unitarity derivation.

Following the remaining steps exactly as above, 
and using $\cK_2^{-1}= q^*\cot \delta(q^*)/(8q^{*2} E)$,
we find
\begin{equation}
\mathcal M_{2L} =  \frac{8 q^* E}{ \cot \delta(q^*) +  \cot \phi(q^*,L) }   \,,
\end{equation}
where
\begin{equation}
 \cot \phi(q^*, L) \equiv  \frac{x }{\pi}  \bigg [   \sum_n -  \PV\int  dn \bigg ]\frac{1  }{  x^2 - n^2 }      =  \frac{x }{\pi}  \frac{\pi  \cot (\pi  x)}{x} = \cot \frac{q^* L}{2}\,,
 \label{eq:1dsum}
\end{equation}
and the second equality follows by noting that the principal-value integral is identically zero and the sum can be evaluated analytically. 
This implies that the finite-volume spectrum is given by all solutions to
\begin{equation}
 \cot \delta(q^*) +  \cot \frac{q^* L}{2} = 0 \,,
\end{equation}
equivalently
\begin{equation}
   e^{2 i \delta(q^*) +i q^* L}    = 1 \,,
\end{equation}
\begin{marginnote}[]
\entry{Quantization condition in $1+1$ dimensions}{}
\end{marginnote}
which is the well-known result for 1+1-dimensional theories~\cite{LuscherWolff}.

\bigskip

These simplified cases illustrate the key steps in the derivation
of the quantization conditions:
\begin{summary}[Key steps]
\begin{enumerate}
\item Demonstrate that power-like finite-volume dependence arises only from $s$-channel diagrams. 
\\[-5pt]
\item Sum contributions from all Feynman diagrams to identify the shift in the finite-volume energies. 
\end{enumerate}
\end{summary}

\subsubsection{General derivation of two-particle quantization condition}
This was first presented in the seminal work of L\"uscher~\cite{Luscher2,Luscher3},
but the approach followed here is more closely based on the derivation of Ref.~\cite{KSS}.  

We begin with the BS equation for the two-particle scattering amplitude
\begin{equation}
\pmb {\mathcal M}_2= \textbf B_2  + \textbf B_2 
\otimes \mathcal I \otimes   \pmb {\mathcal M}_2 \,,
\label{eq:M2}
\end{equation}
shown diagrammatically in {\bf Figure~\ref{fig:1plus2}(c)}.
Here we have introduced a boldface notation in which coordinate dependence, 
 and factors of $i$, are  suppressed, e.g.
$\textbf B_2 \equiv i B_2(P; k'; k)$.
The symbol $\otimes\cI\otimes$ here indicates an integral over the two-particle loop, now in $3+1$ dimensions,
\begin{equation}
\label{eq:BSM2}
 \textbf B_2 \otimes \mathcal I \otimes   \pmb {\mathcal M}_s \equiv 
\frac12 \int \frac{d^4 k}{(2 \pi)^4}   i B_2(P;k'';k')
    \frac{i z(k')}{k'^2 - m^2 + i \epsilon} \frac{i z(P-k')}{(P-k')^2 - m^2 + i \epsilon} i \mathcal M_2(P;k';k) \,,
\end{equation}
where $z(k)/(k^2 - m^2+\ieps)$ is the fully dressed propagator, normalized so that
$z=1$ on shell.

To determine the finite-volume spectrum, we again use $\cM_{2L}$, although, in fact,
any finite-volume correlator would suffice [a different choice was made in Ref.~\cite{KSS}].
We now use the result described above that the 
finite- and infinite-volume versions of the BS kernel differ by terms scaling as $e^{- m L}$
if $m < E^* < 3m$ (or $0 < E^* < 4m$ is there is a $\mathbb Z_2$ symmetry separating
even- and odd-particle number sectors).
This result allows us to write the BS equation for the finite-volume correlator 
[shown in {\bf Figure~\ref{fig:1plus2}(e)}]:
\begin{align}
\pmb {\mathcal M}_{2L}&= \textbf B_2  +
 \textbf B_2 \otimes \mathcal S \otimes   \pmb {\mathcal M}_{2L} \,,
\label{eq:MLinitdecom}
\\
 \textbf B_2 \otimes \mathcal S \otimes   \pmb {\mathcal M}_{2L} 
&\equiv \frac12\int \frac{d k'^0}{2 \pi} \frac{1}{L^3} \sum_{\vec k'} 
 i B_2(P;k'';k')    \frac{i z(k')}{k'^2 - m^2 + i \epsilon} 
\frac{i z(P-k')}{(P-k')^2 - m^2 + i \epsilon} i \mathcal M_{2L}(P;k';k) \,.
\label{eq:BSML}
\end{align}
Here the finite-volume momentum $\vec k'$ is summed over the values $2\pi \vec n/L$,
with $\vec n$ a vector of integers.
We next replace the sum with an integral and a sum-integral difference. All power-like
$L$ dependence lies in the latter quantity. As in the $1+1$-dimensional analysis above,
and as shown in detail in Refs.~\cite{KSS,HS1}, the sum-integral difference picks out the
on shell values of the quantities on either side of $\cS$.
Specifically, it is found that
\begin{align}
\label{eq:SisIplusF}
 \textbf B_2 \otimes \mathcal S \otimes   \pmb {\mathcal M}_{2L} &=  
\textbf B_2 \otimes \mathcal I \otimes   \pmb {\mathcal M}_{2L} 
+  \textbf B_2 \textbf F_2^\ieps     \pmb {\mathcal M}_{2L} 
\\
  \textbf B_2 \textbf F_2^\ieps     \pmb {\mathcal M}_{2L}  
& \equiv  i B_{2;\ell''' m'''; \ell'' m'' }(P) \, i F_{2;\ell'' m''; \ell' m'}^{i\epsilon}(P,L) 
\, i \mathcal M_{2L; \ell' m'; \ell m}(P) \,, 
\\
 i F_{2;\ell'' m''; \ell' m'}^{i\epsilon}(P,L)   & \equiv   
\frac{1}{2}  \bigg[ \frac{1}{L^3} \sum_{\vec k} - \int \frac{d^3 \vec k}{(2 \pi)^3}\bigg]
\frac{i \mathcal Y_{\ell'' m''}(\vec k^*)  \mathcal Y^*_{\ell' m'}(\vec k^*)  }
{2 \omega_{k} 2 \omega_{  P - k} \ 
(E -  \omega_{  \ell} - \omega_{  P -   k} + i \epsilon)}   \,.
 \label{eq:Fdef}
  \end{align}
  \begin{marginnote}[]
\entry{$F_2^\ieps(P,L)$}{Matrix of geometric functions describing $L$ dependence.}
\end{marginnote}
where the product in the second line is now a matrix product with indices given by the
angular momentum in the c.m.~frame of the two on-shell particles.
The quantity $F_2^{i\epsilon}$ is 
 a matrix of geometric functions that encodes how angular momentum-states 
mix due to the reduced symmetry of the finite-volume system.\footnote{%
Both the sum and integral must be UV-regulated. The choice of regulator is
unimportant, however, as different choices lead to results for
$F_2^\ieps$ differing only by exponentially suppressed terms.
$F_2^\ieps$ is related to the generalized zeta functions defined in 
Ref.~\cite{Luscher2,Luscher3}; the explicit relation is given, e.g., 
in Ref.~\cite{HSth}.}
It is the generalization to three-spatial dimensions of the simple one-dimensional
sum evaluated in Eq.~(\ref{eq:1dsum}).
We have introduced $\mathcal Y_{\ell m}(\vec k^*) = 
\sqrt{4 \pi}   (k^*/q^*)^\ell Y_{\ell m}(\hat k^*)$,
where the  prefactor multiplying the spherical harmonic removes 
spurious singularities near $\vec k^* = \vec 0$.

At this stage the derivation proceeds by substituting 
Eq.~(\ref{eq:SisIplusF}) into Eq.~(\ref{eq:MLinitdecom}) and rearranging
\begin{align}
\label{eq:M2LwithF}
\pmb {\mathcal M}_{2L} & = 
\textbf B_2  + \textbf B_2 \big [  \otimes \mathcal I \otimes + \textbf F_2^\ieps \big ]   
\pmb {\mathcal M}_{2L} \,, 
\\
& = \sum_{n=0}^\infty   \textbf B_2 \Big [ \big [  \otimes \mathcal I \otimes + \textbf F_2^\ieps \big ]
    \textbf B_2 \Big ]^n \,, \\
& = \sum_{n=0}^\infty  \Big (\sum_{n'=0}^\infty    
\textbf B _2 \big [ \otimes \mathcal I \otimes     \textbf B_2 \big ]^{n'}    \Big )
 \bigg [ \textbf F_2^\ieps      \Big (\sum_{n'=0}^\infty    \textbf B_2  
\big [ \otimes \mathcal I \otimes     \textbf B_2 \big ]^{n'}    \Big ) \bigg ]^n \,, \\
 & = \sum_{n=0}^\infty  \pmb {\mathcal M}_2
 \big[ \textbf F_2^\ieps      \pmb {\mathcal M}_2  \big ]^n  
=  \pmb {\mathcal M}_2   \frac{1}{1 - \textbf F_2^\ieps  \pmb {\mathcal M}}_2
\label{eq:M2Lfinal} \,.
\end{align}
In the first line we have substituted the identity for $\mathcal S$,
Eq.~(\ref{eq:SisIplusF}). 
Then we have iteratively substituted the expression for $\pmb {\mathcal M}_{2L}$ 
to display all volume dependence. 
In the third line we have regrouped into separate sums over $\mathcal I$ and 
$\textbf F_2^\ieps$. 
In the final step, we have identified the sums involving $\mathcal I$ with $\cM_2$,
using Eq.~(\ref{eq:M2}). Alternatively, one can formally solve Eq.~(\ref{eq:M2LwithF}) directly for $ \pmb {\mathcal M}_{2L}^{-1} $
\begin{equation}
 {\pmb{\mathcal M}}_{2L}^{-1}   = \big [ \textbf B_2^{-1} - \otimes \, \mathcal I \otimes \! \big ] -  \textbf F_2^\ieps \,,
\end{equation}
and identify the square-bracketed contribution as ${\pmb {\mathcal M}}_2^{-1}$. But since the final step is difficult to justify without an all-orders expansion in $\textbf B_2$, it is not clear to us that this more direct line adds anything to the derivation.

We deduce that, for fixed values of $\vec P$ and $L$, the finite-volume energy spectrum in the region $E^* < 3 m$ is given (up to $e^{- mL}$ corrections) by all solutions in $E$ to the quantization condition
 \begin{marginnote}[]
\entry{Two-particle result}{L\"uscher quantization condition for two-particle states.}
\end{marginnote}
\begin{equation}
\det_{\ell' m'; \ell m} \big [\mathcal M_2^{-1}(E^*)  + F_2^\ieps(E, \vec P, L) \big ] = 0\,,
\label{eq:QC2}
\end{equation}
where $\mathcal M_{2;\ell' m'; \ell m} = \delta_{\ell' \ell} \delta_{m'm} \mathcal M_2^{(\ell)}$.
 The extensions to  multiple coupled two-particle channels, 
including different species and particles with spin,  are known, 
as reviewed in Ref.~\cite{Raulreview}.

In the following it will be useful to express the quantization condition in terms of the
K matrix. To this end we introduce a version of $F_2$ in which the PV prescription
is used in the integral in Eq.~(\ref{eq:Fdef}) rather than the $i\epsilon$ prescription.
We denote this simply as $F_2$ without a superscript. It is straightforward so show
[see, e.g., Ref.~\cite{HS1}] that the quantization condition (\ref{eq:QC2}) can
be exactly rewritten as
 \begin{marginnote}[]
\entry{Two-particle quantization condition}{K-matrix form}
\end{marginnote}
\begin{equation}
\det_{\ell' m'; \ell m} \big [\mathcal K_2^{-1}(E^*)  + F_2(E, \vec P, L) \big ] = 0\,,
\label{eq:QC2K}
\end{equation}
where $\mathcal K_{2;\ell' m'; \ell m} = \delta_{\ell' \ell} \delta_{m'm} \mathcal K_2^{(\ell)}$.

\subsection{Three-particle quantization condition in the RFT approach}
\label{sec:QC3}

We now turn to the derivation of the three-particle quantization condition, presented in Refs.~\cite{HS1,HS2}. 

As is discussed in detail in those publications, the quantization condition depends on an intermediate quantity, referred to as a divergence-free three-particle K matrix and denoted by $\Kdf$. This is a non-standard object that encodes the short-distance part of the the three-particle scattering amplitude. It has the same degrees of freedom as the standard scattering amplitude but should be easier to extract from the finite-volume spectrum, since it is a smooth, real function. Most importantly, its relation to the standard $\textbf 3 \to \textbf 3$ scattering amplitude, $\mathcal M_3$, is known \cite{HS2} and depends only on the on-shell $\textbf 2 \to \textbf 2$ scattering amplitude. Thus, the envisioned work-flow is summarized by
\begin{equation}
E_n(L) \ \   \Longrightarrow \ \  \Kdf, \mathcal M_2  \ \ \Longrightarrow \ \  \mathcal M_3 \,.
\end{equation}
 \begin{marginnote}[]
\entry{Basic work-flow}{Relating the finite-volume energies to $\mathcal M_3$.}
\end{marginnote}

In this section we sketch the derivation of the quantization condition for three identical relativistic scalar particles. Following Refs.~\cite{HS1,HS2}, we restrict attention to  theories
 with a $\mathbb Z_2$ symmetry that decouples the even- and odd-particle-number sectors. In addition we assume that the two-particle K matrices have no poles in the kinematic region of interest.
An example of such a theory is the $3\pi^+$ system in QCD in the isospin limit, where
$G$-parity plays the role of the $\mathbb Z_2$ symmetry, and the $2\pi^+$ subsystem is not resonant.
The generalizations to include $\textbf 2 \to \textbf 3$ scattering and to describe systems with sub-channel resonances (generating K-matrix poles) have been worked out in Refs.~\cite{BHS2to3} and \cite{BHSK2poles}, respectively.

The derivation we present is a simplified version of that given in Ref.~\cite{HS1}, based
on the improved approach introduced in Ref.~\cite{BHSK2poles}.
We focus on the important steps, pointing the reader to Refs.~\cite{HS1,BHSK2poles}
for detailed justifications.
We begin in Sec.~\ref{sec:prelim} by setting up  the strategy of the derivation,
based on a skeleton-expansion of a finite-volume correlation function, $C_L$
In Sec.~\ref{sec:decom} we summarize how the $L$ dependence is isolated for 
diagrams of all topologies. 
We then, in Sec.~\ref{sec:combine},
combine results to reach a closed form for $C_L$, from which immediately follows
the quantization condition in terms of $\Kdf$.
Finally, in Sec.~\ref{sec:KtoM} we review the relation between $\Kdf$ and $\mathcal M_3$.

\subsubsection{Preliminaries\label{sec:prelim}}

As in the two-particle case, the derivation presented here is carried out to all perturbative orders in a generic, relativistic quantum field theory. By ``generic'' we mean that no assumptions about the vertices or power-counting scheme are required. 

Consider the finite-volume correlation function
\begin{equation}
C_L(E, \vec P) \equiv \int_L d^4 x \, e^{- i E t +  i \vec P \cdot \vec x} \langle \sigma(x) \sigma^\dagger(0) \rangle_L  \,,
\end{equation}
\begin{marginnote}[]
\entry{Finite-volume correlator}{Poles give spectrum}
\end{marginnote}
where $\sigma^\dagger(0)$ has odd-particle quantum numbers. In the region $m < E^* < 5m$ (where $m$ is the physical particle mass and $E^* = \sqrt{E^2 - \vec P^2}$), all power-like $L$-dependence arises from intermediate three-particle states.

\begin{figure}
\begin{center}
\includegraphics[width=1\textwidth]{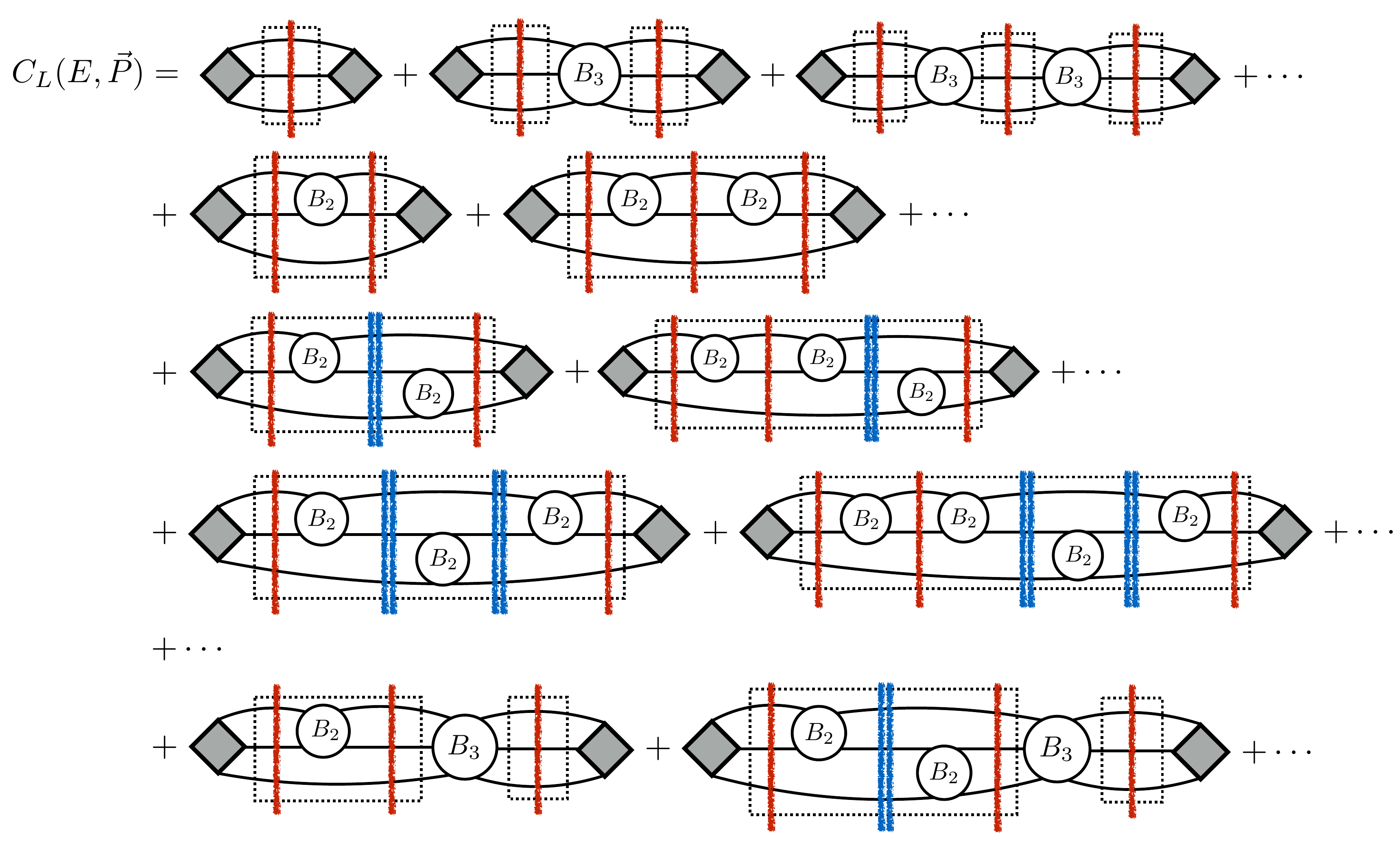}
\caption{Skeleton expansion for finite-volume correlator $C_L$.
Filled triangles indicate the vacuum-to-three-particle matrix elements of the creation
and annihilation operators, called $i\sigma^{\dagger *}$ and $i\sigma^*$, respectively,
in the text. Solid lines are fully dressed propagators with unit on-shell residue.
Unfilled circles are infinite-volume BS kernels $B_2$ and $B_3$, as indicated.
Dashed rectangles enclose three-momenta that are summed.
In the subsequent analysis, sums are replaced by integrals plus sum-integral differences
in places denoted by the red vertical lines, and this leads to so-called $F$ cuts.
Where the interacting pair switches, there is a contribution in which the sum is
kept explicit, and this leads to the geometric quantity $G$. The places where this occurs
are indicated by double blue vertical lines.}
\label{fig:3plus4}
\end{center}
\end{figure}

With this in mind, in Ref.~\cite{HS1} we construct a skeleton expansion in terms of 
BS kernels and fully dressed propagators, as shown in {\bf Figure~\ref{fig:3plus4}}.
The expansion is entirely motivated by the goal of displaying all important $L$-dependence, equivalently all on-shell or long-distance intermediate states. The final structure can be expressed in terms of two kernels: $B_2$, already used above, 
and $B_3$, which contains all 3-particle irreducible diagrams in three-to-three scattering (as well as one-particle propagation that introduces no singularities in the kinematic window considered). 
Both $B_2$ and $B_3$ have only exponentially suppressed $L$ dependence, and
thus, 
under the approximation guiding the derivation, can be replaced by their infinite-volume counterparts.

As we explain in more detail in the following, the relevant volume-dependence, generated by sums over $\vec k = 2 \pi \vec n/L$ in the loops of the skeleton expansion, can be expressed in terms of two finite-volume geometric functions, denoted $F$ and $G$. The first is closely related to the zeta-function of Eq.~(\ref{eq:Fdef}) and  is defined in terms of the PV version of
that quantity:\footnote{%
The PV version is used since this leads to a quantization condition involving $\cK_2$
instead of $\cM_2$ [as in Eq.~(\ref{eq:QC2K})], and thus avoids the cusps in $\cM_2$
that can lead to power-law finite-volume dependence~\cite{HS1}.} 
\begin{equation}
F_{k' \ell' m'; k \ell m} (P, L)  \equiv \delta_{k' k} H(\vec k) F_{2; \ell' m, \ell m}(P-k, L) \,.
\label{eq:F}
\end{equation}
\begin{marginnote}[]
\entry{Kinematic function $F$}{Related to $F_2$}
\end{marginnote}
Here we display the indices that all matrices in the three-particle quantization condition
share: $\vec k$, $\ell$, and $m$. These can be understood by separating the three particles
into a dimer [called ``scattering pair'' in Ref.~\cite{HS1}] and a spectator.\footnote{%
The term ``dimer'' can potentially lead to confusion as different definitions appear in the literature. In this work a dimer is simply a pair of particles projected to a definite orbital angular momentum.}
The spectator is constrained to have one of the discrete finite-volume momenta, $\vec k = 2 \pi \vec n/L$,
while the dimer is decomposed into angular momentum states in its c.m.~frame
(as was done in the two-particle quantization condition), leading to the $\ell m$ indices.
The $\delta_{k'k}$ in Eq.~(\ref{eq:F}) thus represents a situation in which the
spectator does not interact. The argument of $F_2$ gives the four-momentum flowing
through the dimer, $(P-k)^\mu = (E - \omega_k, \vec P -  \vec k)$. 

A new feature in Eq.~(\ref{eq:F}) is the appearance of the cutoff function $H(\vec k)$.
For fixed $P$, as $|\vec k|$ increases, the dimer c.m.~energy,
\begin{equation}
E_{2,k}^* = \sqrt{(E-\omega_k)^2 - (\vec P-\vec k)^2}\,,
\label{eq:E2kstar}
\end{equation}
passes below threshold $E_{2,k}^*=2m$ and eventually drops to zero.
For technical reasons explained in Ref.~\cite{HS1} the formalism requires that
$E_{2,k}^{*2} \ge 0$. The cutoff function $H(\vec k)$ accomplishes this by smoothly
varying from unity at and above threshold to zero when $E_{2,k}^*=0$.
An explicit example of the cutoff is given in Ref.~\cite{HS1}, but will not be needed here.
One can think of it as a soft cutoff at $|\vec k|\sim m$.

The presence of $H(\vec k)$ implies that the index $\vec k$ runs over a finite number
of values. The full matrix space remains infinite dimensional, however, due to
the angular-momentum degrees of freedom. 
If these are truncated, as is common practice in the two-particle sector, 
then $F$ reduces to a finite-dimensional matrix.

The locations in the diagrams of the skeleton expansion at which an $F$ appears is
shown in {\bf Figure~\ref{fig:3plus4}}.
A quantization condition based only on $F$ would predict three-particle energies of the form $E_n(L) = \omega_k + E_2(L, \vec P - \vec k)$ where the second term is an interacting two-particle level with the indicated momentum. Of course, for three identical particles, this cannot be the correct spectrum. It properly encodes the interactions of two, but neglects the third which enters as a non-interacting constituent, albeit with the proper relativistic energy. This motivates the appearance of the second geometric function, $G$, that encodes the volume effects of an exchange in the scattering pair. The explicit definition is\footnote{%
In Refs.~\cite{HS1,HS2} a slightly different form of $G$ was used, in which the pole
term has the form of that in the definition of $F_2$, Eq.~(\ref{eq:Fdef}).
The difference between the two pole terms is nonsingular, 
and can be absorbed by a change in the definition of $\Kdf$. It was
later realized that using the relativistic form shown here leads to a $\Kdf$ that is
relativistically invariant~\cite{BHS2to3}.}
\begin{equation}
G_{p\,\ell' m';k\,\ell m} \equiv \left(\frac{k^*}{q_p^*}\right)^{\ell'}
\frac{4\pi Y_{\ell'm'}(\hat k^*)H(\vec k) H(\vec p) Y^*_{\ell m}(\hat p^*)}
{(P-p -k)^2 - m^2} 
\left(\frac{p^*}{q_k^*}\right)^{\ell}
\frac1{2\omega_k L^3}
\,,
\label{eq:G}
\end{equation}
\begin{marginnote}[]
\entry{Kinematic function $G$}{Arises from exchange of scattering pair}
\end{marginnote}
where $\vec k^*$ is $\vec k$ boosted to the dimer c.m.~frame when $\vec p$ is
the spectator momentum, and 
 $q_p^*=\sqrt{E_{2,p}^{*2}/4 -  m^2}$ is the momentum of each of the two dimer constituents when $\vec p$ is the spectator momentum and all particles are on shell. 
The same definitions hold for $\vec p^{\,*}$ and $q^*_k$, with the roles $\vec k$ and $\vec p$ exchanged.
Examples of the locations in skeleton-expansion diagrams that lead to factors of $G$ are
shown in {\bf Figure~\ref{fig:3plus4}}. 

 At this stage we are left to explain three more-technical aspects of the notation. First, at various intermediate stages we consider sums over three-to-three diagrams in which an incoming or outgoing particle is singled-out by the property of being \emph{unscattered} by the outermost two-to-two insertion. We use the superscript $(u)$ to denote such unsymmetrized quantities. Given that we consider identical particles, the final result cannot (and, as we prove in Ref.~\cite{HS1}, does not) depend on these incomplete objects. Second, we find it convenient to introduce a bold-faced notation for the BS kernels as well as other building blocks defined via partial diagrammatic sums. This simply denotes that factors of $i$ and $1/(2 \omega L^3)$ have been absorbed to simplify expressions, 
  and is taken over from Ref.~\cite{BHSK2poles}.
 Third, it is convenient to begin the derivation by studying only the part of $C_L(E, \vec P)$ that survives when we send $B_3 \to 0$. This is indicated by a $[B_2]$ superscript, $C_L \to C_L^{[B_2]}$.

With these preliminaries established, we are now ready to jump into the derivation.

\subsubsection{Decomposing to the level of $\bKdf$\label{sec:decom}} 
We begin with Eq.~(174) of Ref.~\cite{HS1}, expressed in boldface notation
as in Eq.~(49) of Ref.~\cite{BHSK2poles},
\begin{align}
C^{[B_2]}_{L}  &= 
 C_{L,0F} 
- \frac23 \bSig  \bF  \bSigD
+ 
\bA'^{(u)}_{L,3} 
\bF_{33}^{(0)}
\sum_{n=0}^\infty 
\left( \bKLth^{(u,u)}  
       \bF_{33}^{(0)} \right)^n  
\bA^{(u)}_{L,3} \,,
\label{eq:Cno3MLres}
\\
\bF^{(0)}_{33} & \equiv \bF \frac{1}{1 - \bK \bF} 
\,. 
\label{eq:F33zerodef}
\end{align}
 \begin{marginnote}[]
\entry{Initial decomposition}{Factorizing $L$-dependence between adjacent $\bK$ factors.}
\end{marginnote}
To obtain this result we have applied
the method that led to the two-particle quantization condition
for diagrams in which adjacent factors of $B_2$ are attached to the same dimer.
This explains the appearance of $\bK \equiv i [2 \omega L^3] \mathcal K_2$, 
which is diagonal in spectator indices,\footnote{%
For technical reasons, in the subthreshold region ($E^*_{2,k} < 2m$), $\cK_2$
is defined to includes a smooth interpolation from the sub-threshold K matrix 
($\propto [\,q^* \cot \delta]^{-1}$) to the sub-threshold scattering amplitude 
($\propto [\,q^* \cot \delta + |q^*|]^{-1}$).}
as well as $\bF \equiv i F/(2 \omega L^3)$.

\begin{figure}
\begin{center}
\includegraphics[width=1\textwidth]{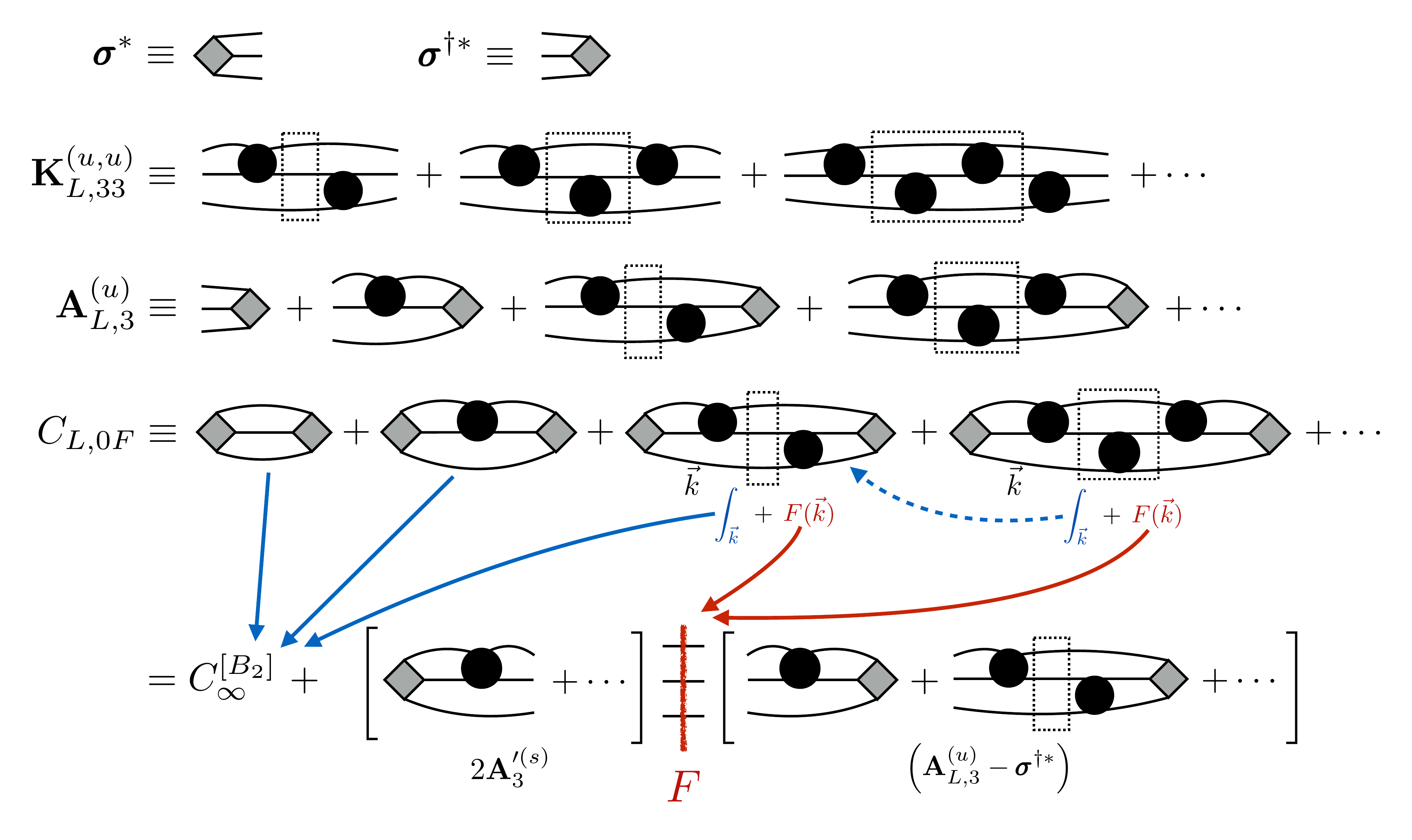}
\caption{Definitions of quantities appearing in Eq.~(\ref{eq:Cno3MLres}), using
notation from previous figures. Filled circles represent two-particle K matrices,
$\bK$. The quantity $\bA'^{(u)}_{L,3}$ is given by the horizontal reflection 
of the diagrams for $\bA^{(u)}_{L,3}$.
Also shown is a sketch of the derivation of the decomposition of
$C_{L,0F}$ given in Eq.~(\ref{eq:CL0Ffinal}), in which the sum over the left-most
summed loop integral is replaced by an integral plus sum-integral difference, the
latter leading to a factor of $F$. For the integral term, if there are further sums remaining,
then the procedure is repeated, as indicated by the dashed blue arrow.
}
\label{fig:5plus6}
\end{center}
\end{figure}

The other quantities in Eq.~(\ref{eq:Cno3MLres}) involve diagrams
in which particles pairwise scatter via $\bK$, with the remaining loops (containing exchange propagators) involving sums over the finite-volume momenta. 
These quantities are finite-volume objects, as indicated
by the subscript $L$. They are shown {\bf Figure~\ref{fig:5plus6}}.
Specifically,  $\bKLth^{(u,u)}$ is sum of all fully connected three-to-three scattering 
diagrams involving alternating pairwise scattering,
while  $\bA^{(u)}_{L,3}$, $\bA'^{(u)}_{L,3} $ and $C_{L,0F}$  are defined similarly,
but with additional endcap factors $\bSigD$ and $\bSig$.

The remaining task is to decompose the volume dependence of $C_{L,0F}$, $\bA'^{(u)}_{L,3} $, $\bA^{(u)}_{L,3}$ and $\bKLth^{(u,u)}$, expressing each object as matrix products of infinite-volume quantities and finite-volume geometric functions. To this end one can show that
\begin{align}
C_{L,0F}
&=
{C}^{[B_2]}_{\infty }
+
  2 \bA'^{(s)}_3  {\bF} 
(\bA_{L,3}^{(u)} - \bSigD)
\,, \label{eq:CL0Ffinal}
\\
\bA'^{(u)}_{L,3}
& =
\bA'^{(u)}_3
+
  2 \bA'^{(s)}_3  \bF 
\left(  \bKLth^{(u,u)}+ \bK \right)
\,, 
\label{eq:ApL3final} \\
\bA^{(u)}_{L,3}
& =
\bA^{(u)}_3
+
\left( \bKLth^{(u,u)} + \bK \right)
  \bF \, 2 \bA^{(s)}_3 
\,,
\label{eq:AL3final} 
\end{align}
 \begin{marginnote}[]
\entry{Second decomposition}{Reducing all terms to $\bKLth^{(u,u)}$.}
\end{marginnote}
thereby reducing all unfactorized $L$-dependence to that of $\bKLth^{(u,u)}$. 
The derivation of the result for $C_{L,0F}$ is sketched in {\bf Figure~\ref{fig:5plus6}};
those for the other quantities is similar.
The idea is to move from left to right, 
replacing a given momentum sum with an integral plus a sum-integral difference. 
The term with the difference leads to a factor of $\bF$ whereas that with the integral
 is further decomposed, by applying the same prescription to the next sum in the chain. In this way all contributions are cast into the same form: an $\bF$-cut with an infinite-volume expression to the left and remaining $L$ dependence to the right.

At this stage it only remains to decompose $ \bKLth^{(u,u)} $. One finds 
\begin{equation}
 \bKLth^{(u,u)}  
= 
 \bKLth^{(0)}
+
[1+ \bT \bG]       \bKdfth^{(u,u)}  \frac1{1 - \bGK  \bKdfth^{(u,u)} }   [1+ \bG \bT ] \,,
\label{eq:KLmatrix3}
\end{equation}
 \begin{marginnote}[]
\entry{Final decomposition}{Displaying all $L$ dependence as products of geometric functions.}
\end{marginnote}
where
\begin{gather} 
\label{eq:KL330}
\bKLth^{(0)}  \equiv 
 \frac1{1  - \bK \bG } \bK \bG \bK \,,\qquad
 \bT  \equiv \bK \frac1{1- \bG \bK} \,, \qquad 
  \bGK  \equiv   \frac1{1- \bG \bK} \bG \,.
\end{gather}
This decomposition, explained in Ref.~\cite{BHSK2poles}, leads to the first appearance
of the second geometric function $\bG =i(2\omega L^3)^{-1} G$, as well of the
intermediate infinite-volume quantity $\bKdfth^{(u,u)}= i\Kdf^{(u,u)}$.
Here the subscript ``df'' stands for divergence free and indicates that kinematic singularities present in the three-to-three scattering amplitude are absent from this quantity.  

\subsubsection{Combining and symmetrizing\label{sec:combine}}

We now substitute all decompositions into Eq.~(\ref{eq:Cno3MLres}) and organize terms by the number $\bKdfth$ factors that they contain. 

Beginning with the $\bKdf$-independent contributions, note that $ \bKLth^{(u,u)}  \longrightarrow \bKLth^{(0)}$ in the limit of $\bKdfth \to 0$. This can be used to show that the factor appearing between $\bA'^{(u)}_{L,3}$ and $\bA^{(u)}_{L,3}$ in Eq.~(\ref{eq:Cno3MLres}) reduces as
\begin{equation}
\bF_{33}^{(0)}
\sum_{n=0}^\infty 
\left( \bKLth^{(u,u)}  
       \bF_{33}^{(0)} \right)^n \ \ \  
       \xrightarrow{\bKdf\to 0}\ \ \  
  \bF \frac1{1- \bT \bF} \equiv \bZ \,.
\label{eq:Zdef}
\end{equation}
It is then straightforward to reduce the two endcaps as well as  $C_{L,0F}$ to reach
\begin{align}
 \bA'^{(u) }_{L,3} &  \ \ \  
 \xrightarrow{\bKdf\to 0}\ \ \  
\bA'_3 
-  2 \bA'^{(s)}_3 (1- \bF \bT )
\,,
\label{eq:ApL3noKdf_v2}
\\[5pt]
 \bA^{(u)}_{L,3} &  \ \ \  
 \xrightarrow{\bKdf\to 0}\ \ \  
\bA_3
- (1 - \bT \bF ) 2 \bA^{(s)}_3 
\,, 
\label{eq:AL3noKdf_v2} \\[5pt]
C_{L,0F} - {C}^{[B_2] }_{\infty } & \ \ \ 
\xrightarrow{\bKdf\to 0}\ \ \  
  2 \bA'^{(s)}_3   {\bF}  \left ( 
\bA_3-(1-\bT \bF) 2 \bA^{(s)}_3 - \bSigD \right)
  \,. \label{eq:CL0FloKdf}
\end{align}
Here we have expressed the endcaps in terms of $\bA'_3 \equiv \bA'^{(u) }_{3} + \bA'^{(s) }_{3}  + \bA'^{(t) }_{3}  $ and similarly for the unprimed object. This combines the three possible choices of momentum assignment for the external particle that is not attatched to the outermost $\bf 2 \to \bf 2$ insertion. Although we are forced to consider unysmmetrized quantities at various stages of the derivation, it is crucial that the final result should only depend on objects that have exchange symmetry with respect to the momenta of the three identical particles. This is proven explicitly for all systems considered so far in Refs.~\cite{HS1,HS2,BHS2to3,BHSK2poles}.

Substituting the four relations [Eqs.~(\ref{eq:Zdef})-(\ref{eq:CL0FloKdf})] into Eq.~(\ref{eq:Cno3MLres}) leads to a complicated expression. However, the important part is given by the term containing the two symmetrized endcap factors $\bA'_3$ and $\bA_3$:
\begin{equation}
 C_L^{[B_2]} - C_\infty^{[B_2]} \supset \bA'_3 \bZ \bA_3   =   \bA'_3   \bigg [      \bF +       \bF \frac1{1- {\bf K}_2 (\bf F + \bf G)}  {\bf K}_2 \bF \bigg] \bA_3 \,,
 \end{equation}
 where in the equality we have given an alternative, expanded expression for $\bZ$. 
 Remarkably, this simple result almost captures the full volume dependence at leading order in $\bKdfth$. The set of remaining contributions modifies the result in two minor ways. First, many of the additional terms can be proven to have only exponentially-suppressed $L$ dependence and are absorbed into a redefinition of $C_\infty^{[B_2]} $. Second, the one additional volume cut that survives (after significant reshuffling) is a term that corrects the numerical factor multiplying $ \bA'_3 \bF \bA_3$. One finds
\begin{align}
 C_L^{[B_2]} - C_\infty^{[B_2]} &= \bA'_3 \bF_{33} \bA_3 + \mathcal O(\bKdf) \,,
 \\
  \bF_{33} &\equiv    \frac13 \mathbf F  + \mathbf F    \frac1{1- {\bf K}_2 (\bf F + \bf G)}  {\bf K}_2    \mathbf F      \,.
 \end{align}
\begin{marginnote}[]
\entry{Leading order result}{$C_L^{[B_2]} $ at leading order in $\bKdf$.}
\end{marginnote}
As we will see below, $\bF_{33}$, which combines geometric functions 
together with factors of $\cK_2$, is the three-particle analog of $\bF_2$. 
It is the central object entering the three-particle quantization condition. 

 \bigskip
 
We now continue the pattern by considering the next order in $\bKdfth$. For example, from the decomposition of $\bKLth^{(u,u)}$ in Eq.~(\ref{eq:KLmatrix3}) together with the $\bKdf$-independent result [Eq.~(\ref{eq:Zdef})] it is straightforward to show that
%
\begin{align}
\bF_{33}^{(0)}
\sum_{n=0}^\infty 
\left( \bKLth^{(u,u)}  
       \bF_{33}^{(0)} \right)^n 
& = \bZ   + \bZ  (1 + \bT \bG)  \,
     \bKdfth^{(u,u)}    \,
   (1 + \bG  \bT )   \bZ     + \mathcal O(\bKdf^2) \,.
   \label{eq:intermediate}
\end{align}
The next step is to identify the $\mathcal O(\bKdf)$ contributions to $\bA'^{(u) }_{L,3}$, $\bA^{(u)}_{L,3}$ and $C_{L,0F} - {C}^{[B_2] }_{\infty }$ and assemble all terms to identify the corresponding contribution to $C_L^{[B_2]} - C_\infty^{[B_2]} $. However, for the purposes of this review, we think it more instructive to consider
  a single contribution to $C_L^{[B_2]}$, given by sandwiching the $\cO(\bKdf)$ part of
Eq.~(\ref{eq:intermediate}) between symmetrized, infinite-volume endcaps,
\begin{multline}
\bA'_3   \bZ  (1 + \bT \bG)  \,
     \bKdfth^{(u,u)}    \,
   (1 + \bG  \bT )   \bZ \bA_3    =   \\ \bA'_3    \bigg [ \bF + \bF \frac1{1- \bK (\bF + \bG)}  \bK (\bF + \bG)  \bigg ]      \bKdfth^{(u,u)} \bigg [    \bF +   (\bF + \bG) \bK        \frac1{1- (\bF + \bG) \bK }  \bF \bigg ] \bA_3  \,.
\end{multline}

As above, the inclusion of all other terms leads to only minor modifications to this key result. Again various $L$-independent terms are absorbed, not only into $C_\infty^{[B_2]} $ but also into $ \bA'_3$ and $\bA_3$. A new feature that arises here is that the unsymmetrized K matrix, $ \bKdfth^{(u,u)}$, can be replaced with a symmetrized form, up to a correction of the symmetry factor on the isolated $\bF$ term ($\bF \to \bF/3$) and a modification of the cuts multiplying $ \bKdfth^{(u,u)}$. The upshot is that the square-bracketed factors are replaced with $\bF_{33}$, the same three-particle volume cut that appeared at the previous order:
\begin{equation}
 C_L^{[B_2]} - C_\infty^{[B_2]} = \bA'_3 \bF_{33} \bA_3 +  \bA'_3 \bF_{33} \bKdfth^{[B_2]} \bF_{33} \bA_3 + \mathcal O(\bKdf^2) \,.
 \end{equation}
  \begin{marginnote}[]
\entry{NLO result}{$C_L^{[B_2]} $ at next-to-leading order in $\bKdf$.}
\end{marginnote}

At this stage, we see the structure emerging and can assign a physical interpretation to the pattern. Within the finite-volume correlator, the three-particle state is created with an insertion of $ \bA_3 $. This is equal to a matrix element of $\sigma^\dagger$ and measures the probability amplitude to create a three-particle state from the vacuum.\footnote{%
Strictly speaking this does not hold because of the removal of singular long-distance contributions. To recover the standard matrix element these have to be put back in, following an approach analogous to the relation between $\mathcal K_{\text{df},3}$ and $\mathcal M_3$ discussed in Sec.~\ref{sec:KtoM} below.} 
The three particles then propagate and rescatter in the box, leading to a pattern of Feynman diagrams that translates into a set of nested geometric series. 

We organize these in powers of the short-distance three-to-three interaction, $\bKdfth $. Then at leading order the volume affects arise due to a single insertion of $\bF_{33}$. This, in turn, is equal to a factor of $\bF$ together with a sum over all terns of the form $\bF \bK \bF \bK \bG \cdots \bF \bK \bF$ where either $\bF$ or $\bG$ can appear between any two factors of $\bK$. The sum over all such terms represents the fact that any particle pair can scatter (inducing a factor of $\bK$) and then propagate between consecutive re-scattering events (giving $\bF$), or alternatively exchange the scattering pair (leading to $\bG$). Finally the term that is linear in $\bKdfth$ is governed by this same structure, together with the observation that a short-distance three-particle interaction may arise anywhere in the series of two-particle scattering events.

From the physical intuition it is perhaps not so surprising that the same pattern continues to all orders in $\bKdf$. This is proven explicitly in Refs.~\cite{HS1} and \cite{BHSK2poles}. In addition, the inclusion of the three-particle BS kernels turns out to only modify the definition of the divergence-free K matrix. This can be accommodated by 
replacing $\bKdfth^{[B_2]}$ with $\bKdfth$. 
 Putting all this together leads to the main result of Ref.~\cite{HS1}
 \begin{equation}
 C_L^{ } - C_\infty^{ } =  \sum_{n=0}^\infty \bA'_3 \bF_{33} \Big [    \bKdfth^{ } \bF_{33}     \Big]^n \bA_3 = \bA'_3 \bF_{33} \frac{1}{1-  \bKdfth^{ } \bF_{33}    } \bA_3 \,.
 \end{equation}
 
Thus the poles in the finite-volume correlator occur whenever the matrix appearing between $ \bA'_3$ and $ \bA_3$ has a divergent eigenvalue. We deduce that, for fixed values of $\vec P$ and $L$, the finite-volume energy spectrum in the region $m < E^* < 5 m$ is given (up to $e^{- mL}$ corrections) by all solutions in $E$ to the quantization condition
\begin{equation}
\det_{\vec k' \ell' m'; \vec k \ell m} \big [\mathcal K_{\text{df},3}(E^*)  +  F_3(E, \vec P, L)^{-1} \big ] = 0 \,.
\label{eq:QC}
\end{equation}
 \begin{marginnote}[]
\entry{Three-particle quantization condition}{RFT approach}
\end{marginnote}
Here we have returned to the non-bold notation,
 using $\bF_{33} = i F_3$ and $\bKdfth=i\Kdf$,
 that we will use for the remainder of the discussion.

 \subsubsection{Relating $\mathcal K_{\mathrm{df},3}$ to $\mathcal M_{3}$\label{sec:KtoM}}

In the previous sections we have related $\mathcal K_{\text{df},3}(E^*) $ to the finite-volume energy spectrum. This object can, in principle, be constrained by calculating finite-volume energies, for example from the Euclidean-time decay of correlators  calculated
numerically using LQCD, and then applying Eq.~(\ref{eq:QC}). Of course, this is only of interest if $\mathcal K_{\text{df},3}(E^*) $  can be related to infinite-volume observables.

Indeed, as was shown in Ref.~\cite{HS2}, this modified K matrix is related to the three-to-three scattering amplitude via an integral equation that depends only on known functions as well as the on-shell two-particle scattering amplitude. The relation has a number of desirable features. The equations are defined at fixed energy, meaning that $\mathcal K_{\text{df},3}(E^*) $ is only required for $E^*$ where the physical scattering amplitude is to be determined. In addition once the quantization condition is fixed, there is no ambiguity or scheme dependence in the relation between $\mathcal K_{\text{df},3}$ and $\mathcal M_3$. Finally the $\mathcal K_{\text{df},3}$ to $\mathcal M_3$ relation manifestly encodes the complicated unitarity constraints of the three-particle scattering amplitude~\cite{BHSS}.

The relation is derived in a slightly round-about way. In particular we show in Ref.~\cite{HS2} that one can define an alternative finite-volume correlation function, $\mathcal M_{3L}$, that becomes the physical scattering amplitude in a carefully constructed $L \to \infty$ limit. 
This new correlator differs from $C_L$ only in the interpolating fields. As a result it admits a similar skeleton expansion and a similar decomposition into geometric functions. In  Ref.~\cite{HS2} we show that
\begin{equation}
  \mathcal M_{3L}[\mathcal K_{\text{df},3}, \mathcal K_2] \equiv \mathcal S \bigg[   \mathcal D_L^{(u,u)}    -   \mathcal L_L^{(u)}       \frac{1}{1 +   \mathcal K_{\text{df},3}    F_3 }        \mathcal K_{\text{df},3}    \mathcal R_L^{(u)}  \bigg ] \,,
  \label{eq:M3Lres}
 \end{equation}
 where $ \mathcal D_L^{(u,u)} $, $ \mathcal L_L^{(u)} $ and $\mathcal R_L^{(u)} $ are 
 known explicitly, and are closely related to $F_3$, while $\cS$ indicates symmetrization.

 As an aside, we observe from Eq.~(\ref{eq:M3Lres})
 that we can use $\cM_{3L}$ instead of $C_L$ in order
 to derive the quantization condition, as its poles lie at the same positions.
 Indeed, the unsymmetrized quantity in square braces is very similar to the
 dimer-particle scattering amplitude used in the alternative approaches to deriving
 the quantization condition discussed below.

We now obtain $\cM_3$ by taking a judiciously chosen $L\to\infty$ limit of $\cM_{3L}$.
This requires that factors of $i\epsilon$ in the pole terms be first put back in, so the
$L\to\infty$ limit is well defined.
Thus we have
\begin{equation}
 \mathcal M_{3}(\mathcal K_{\text{df},3}, \mathcal K_2)  = \lim_{\epsilon \to 0} \lim_{L \to \infty} \mathcal M_{3,L}(\mathcal K_{\text{df},3}, \mathcal K_2)  \,,
\end{equation}
which gives a set of purely infinite-volume integral equations, given
explicitly in Ref.~\cite{HS2}.
These map the real, short-distance three-body quantity $\Kdf$ to the complex, unitary, relativistic three-to-three scattering amplitude.

 \subsection{Truncating the quantization condition}
 \label{sec:trunc}
 
The quantization condition Eq.~(\ref{eq:QC})  is a formal result, because the determinant runs over an infinite-dimensional matrix space. This is exactly as in the two-particle case and, in direct analogy to that case, truncating $\mathcal K_2$ and $\mathcal K_{\text{df},3}$ to vanish above some $\ell_{\text{max}}$ is sufficient to reduce all matrices in the quantization condition to have finite dimension~\cite{HS1}. Specifically, the matrices then have dimension $[(2 \ell_{\text{max}} + 1) \times N_{\vec k}]$ where $N_{\vec k}$ is the number of finite-volume momenta, $\vec k$, for which the cutoff function, $H(\vec k)$, is non-zero.
 
 The simplest approximation is to take $\ell_{\text{max}}=0$: the $s$-wave approximation.
 This is the approximation used from the beginning in the
 alternative approaches described below.
 It assumes that $\mathcal K_2$ and the dimer within $\mathcal K_{\text{df},3}$
 are both dominated by $s$-wave interactions. 
 Given this truncation, $\mathcal K_2$ reduces to a single function of the two-particle CM
 energy, whereas $\mathcal K_{\text{df},3}$ retains dependence on the
 spectator momenta $\vec k'^*$ and $\vec k^*$.
 This is the form of the RFT quantization condition that is used in the
 comparison with the results from alternate approaches;
 it is given explicitly in Eq.~(\ref{eq:F3s}) below.
 
The $s$-wave approximation is well motivated at energies close to the three-particle
 threshold, since higher waves are then suppressed by factors of  $(q_k^{*})^{\ell}$.
To study this systematically, one expands $\cK_2$ and $\Kdf$ about
 threshold in powers of $s-9m^2$ and related quantities. 
 For $\cK_2$ this leads to the effective range expansion, with
 the leading term being the scattering length. The expansion of $\Kdf$ is
 constrained because it is relativistically invariant,
 symmetric under initial and final particle interchange, 
 and time reversal invariant~\cite{BHS2to3}.
  Using these symmetries, one can show that the first two terms in the expansion
 are not only pure $s$-wave but  also isotropic,  i.e. independent of the spectator momenta~\cite{BHS2to3,BBHRS,BRS}.
 At third order one must include $d$-wave contributions in both $\cK_2$ and $\Kdf$~\cite{BRS}.
 
 The $s$-wave approximation plus isotropic $\Kdf$ is referred to as the isotropic approximation.
 In this most extreme truncation $\Kdf$ depends only on $E^*$,
 so we can write 
 \begin{equation}
 \mathcal K_{\text{df},3} = \delta_{\ell,0} \delta_{m,0} \vert 1 \rangle  \mathcal K^{\text{iso}}_{\text{df},3}(E^*) \langle 1 \vert \,,
 \end{equation}
 where $\vert 1 \rangle $ is an unnormalized vector with unit entry for every active
 spectator momentum.
If one further restricts to the $A_1$ irreducible representation (irrep) of the cubic group,
then, as shown in Ref.~\cite{HS1}, all nontrivial solutions to the quantization condition
satisfy
 \begin{equation}
 \mathcal K^{\text{iso}}_{\text{df},3}(E^*)   = -1/F_3^{\text{iso}}(E, \vec P, L) \,,\quad
 F_3^{\text{iso}} \equiv \langle 1 \vert F_{3,s} \vert 1 \rangle\,,
 \end{equation}
 where $F_{3,s}$ is the form of $F_3$ after $s$-wave truncation [see Eq.~(\ref{eq:F3s})].
Thus, in this approximation, we recover a one-to-one correspondence between finite-volume energy levels and the value of $\mathcal K^{\text{iso}}_{\text{df},3}(E^*) $.
This approximation has been studied numerically in Ref.~\cite{BHSnum},
and some of the results are shown in Sec.~\ref{sec:num} below.

 \subsection{Analytic investigations of the RFT quantization condition}
 \label{sec:anal}
 
 In this subsection we describe two analytic investigations of the three-particle quantization
 condition, Eq.~(\ref{eq:QC}). The main aim is to check the formalism by comparing
 to known results in special limits, but a side-benefit is that some new analytic
 results are obtained.
 
 \subsubsection{$1/L$ expansion}
 \label{sec:threxp}
 Without interactions, the lowest energy state of three particles with $\vec P=0$
 has energy $3m$. Turning on interactions, this level will shift to $E_0(L)$.
 For sufficiently large $L$, $E_0(L)-3m$ can be expanded in powers of $1/L$,
 the so-called threshold expansion.
The leading term scales as $1/L^3$, corresponding to the probability of two particles
to overlap, while three-particle interactions enter first at $1/L^6$.
This expansion was worked out previously, using NRQM, up to 
$\cO(1/L^7)$~\cite{Detmold,Tan,Detmold2}.

We determined the threshold expansion, starting from Eq.~(\ref{eq:QC}), in Ref.~\cite{HSth}.
We found\footnote{%
In Ref.~\cite{HSth}
we write $ \mathcal C_F+ \mathcal C_4+\mathcal C_5$ in place of $\mathcal C'$.}
\begin{multline}
E_0(L) = 3m + \frac{12 \pi a}{m L^3} \bigg \{ 1 - \mathcal I \frac{a}{\pi L} +   \left(\frac{a}{\pi L}\right)^{\!2} \left({\cal I}^2+{\cal J}\right) + \frac{64\pi^2 a^2 }{mL^3} \mathcal C_3+\frac{3\pi a}{m^2 L^3} + \frac{6\pi ra^2}{L^3} \\
+  \left(\frac{a}{\pi L}\right)^{\!3}
\bigg (\! -{\cal I}^3 + {\cal I}{\cal J} + 15 {\cal K} 
+\frac{16 \pi^3}{3} (3\sqrt 3 - 4\pi) \, \log \! \bigg (\frac{mL}{2\pi} \bigg ) + \mathcal C' \bigg ) \bigg \} - \frac{ \mathcal M_{3,\mathrm{thr}}}{48 m^3 L^6} + \mathcal O \! \left(\frac{1}{L^7} \right) \,,
\label{eq:thresh_res}
\end{multline}
where we have introduced the following geometric constants: $\mathcal I=-8.914$, $\mathcal J=16.532$, $\mathcal K=8.402$, $\mathcal C_3 = -0.05806$, $\mathcal C' = 2052$,
whose origin is explained in Ref.~\cite{HSth}.
The scattering parameters that enter here are the scattering length, $a$, the effective range, $r$, and the threshold three-to-three scattering amplitude $\mathcal M_{3,\mathrm{thr}}$. The latter has a somewhat subtle definition, due to the fact that the full scattering amplitude diverges at threshold and thus requires a subtraction. Again see Ref.~\cite{HSth} for details.
Note that the leading term is three times that found in Eq.~(\ref{eq:HYth}) for the
two-particle system, as expected since there are three pairs of particles that can interact.

The result of Eq.~\eqref{eq:thresh_res} agrees with the
results of Refs.~\cite{Detmold,Tan} for the terms through $\mathcal O(1/L^5)$.
This provides a strong check on the formalism.
Relativistic effects enter at $\mathcal O(1/L^6)$ and here no general check is available.
However the $\log L$ term is universal and its coefficient also agrees with earlier work. 
To check the remaining $1/L^6$ terms, in Refs.~\cite{HSPT,SPT} we calculated
$E_0(L)$  in $\lambda \phi^4$ theory up to $\cO(\lambda^4)$, 
finding complete agreement with Eq.~(\ref{eq:thresh_res}).
Working at this order checks all the terms entering at $1/L^6$.

We also note that, for weakly interacting systems, the threshold expansion might
provide a partial alternative to the full quantization condition. 
By fitting the $L$ dependence of the threshold state
at many volumes, one could in principle extract $\mathcal M_{3,\mathrm{thr}}$ and thereby determine the near-threshold scattering amplitude.
This approach has been followed successfully in $\lambda \phi^4$ theory 
in the recent work of Ref.~\cite{Fernando}.
  
  \subsubsection{Volume-dependence of a three-body bound state}
  Our second analytic result concerns 
the volume-dependence of a non-relativistic three-scalar Efimov bound state~\cite{Efimov}
in the unitarity limit for two-particle interactions, i.e. with $1/a \to 0$.
 This was studied in Ref.~\cite{MRR} using non-relativistic quantum mechanics in a finite-volume. The authors found that, when the infinite volume system contains a bound state, then the lowest lying state in finite volume has energy
 \begin{equation}
 E_B(L) = 3m - \frac{\kappa^2}{m}  + \Delta E(L)  \,,
 \end{equation}
 where $\kappa$ is the binding momentum, and
 \begin{equation}
 \label{eq:MRRresult}
 \Delta E(L) =  c  \vert A \vert^2 \frac{\kappa^2}{m}    
\frac{1}{(\kappa L)^{3/2}} e^{- 2 \kappa L/\sqrt{3} }  + \cdots  \,.
 \end{equation}
 Here the ellipsis indicates terms suppressed by additional factors of $\kappa^2/m^2$ or $1/( \kappa L)$ as well as faster-decaying exponentials. The result depends on $c  \simeq -96.351$, a known geometric constant, and $\vert A \vert^2$. The latter is a normalization correction that arises because the asymptotic wave-function is not a strict solution to the Schr\"odinger equation. It is expected to be close to unity when the pairwise interactions of the theory are well-described by a short-range potential.

The generic relativistic quantization condition presented above,
Eq.~(\ref{eq:QC}),
holds for systems with a three-particle bound state, 
as long as there are no bound dimers. 
Indeed, in our numerical implementation in the isotropic approximation, we have found
that the formalism can support such bound states~\cite{BHSnum}.
Assuming the existence of such a state 
(i.e. that there is a subthreshold pole in $\mathcal M_3$), 
in Ref.~\cite{HS3bound} 
we were able to provide another check on the quantization condition 
by reproducing the coefficient, power-law envelope and exponential decay given by Eq.~(\ref{eq:MRRresult}). 
In addition we extended the result to non-zero momentum in the finite-volume frame, and found that the moving-frame energy is shifted according to 
 \begin{equation}
\Delta E (\vec P, L) = f_3 \! \big [\vec n  \big] \, \Delta E(L) + \cdots \,,\qquad
f_3\! \big [{\vec n} \big] = \frac{1}{6} \sum_{\hat s }  e^{i (2\pi/3) \hat s \cdot \vec n }\,,
\label{eq:DEP}
\end{equation}
where $\vec n = L \vec P/(2 \pi)$ and the sum runs over the six unit vectors pointing along the finite-volume axes. 
An interesting corollary is that the leading-order volume shift vanishes for $\vec n = (0,1,1)$.

\section{ALTERNATIVE\  APPROACHES}
\label{sec:alt}

As noted in the Introduction, two other approaches for deriving the three-particle quantization
condition have been used, and in this section we describe them, quote the resulting form for the
quantization condition, and explain in what ways these approaches 
agree with and differ from the generic RFT 
approach described in the previous section. 
In particular, we present new results concerning the analytic relation between the three
approaches in the $s$-wave approximation,
which is the only case that has been worked out explicitly within the new approaches.

We discuss the two other approaches in historical order, beginning with that based
on non-relativistic effective field theory (NREFT), introduced in Refs.~\cite{Akaki1,Akaki2},
and then describing the approach based on extending the form of amplitudes
required to satisfy unitarity to finite volume~\cite{MD1}.
We refer to the latter as the ``finite-volume unitarity'' (FVU) approach.

\subsection{NREFT approach}


NREFT is applicable when the momenta of all particles are in the non-relativistic
domain, $|\vec p \, | \ll m$. In this regime there is no particle creation, which
greatly simplifies the diagrammatic analysis. 
Examples where it is 
directly applicable in a three-particle context include three pions close to
threshold in an isospin-symmetric world (so that G-parity forbids three-to-two transitions)
and three nucleons close to threshold.

The first calculations using NREFT to determine finite-volume properties of three-particle
systems were numerical calculations for the triton~\cite{KH08,KH09} and
Efimov states~\cite{KH10}. This approach was then used in Refs.~\cite{Akaki1,Akaki2}
to develop the three-particle quantization condition, which we review
in the following.

The authors consider a theory of identical scalars (i.e. the same setup as used
in the REFT analysis presented earlier) for which the NREFT Lagrangian is\footnote{%
There are also single-particle operators involving higher-order derivatives, which must
be included perturbatively, along the lines of Ref.~\cite{SavageEFT98}.
Implementation of these terms is underway~\cite{AkakiPrivate}.}
\begin{align}
\cL &= \psi^\dagger \left(i \partial_0 + \frac{\nabla^2}{2m}\right) \psi + \cL_2 + \cL_3\,,
\label{eq:L}
\\
\cL_2 &= - \tfrac12 C_0 \psi^\dagger\psi^\dagger \psi \psi
+ \tfrac14 C_2 
\left(\psi^\dagger\overleftrightarrow{\nabla}^2 \psi^\dagger \psi \psi + \textrm{h.c.}\right)
+ \dots\,,
\label{eq:L2}
\\
\cL_3 &= - \tfrac16 D_0\psi^\dagger\psi^\dagger \psi^\dagger\psi\psi \psi
+ \tfrac1{12} D_2 
\left(\psi^\dagger\psi^\dagger\overleftrightarrow{\nabla}^2 \psi^\dagger \psi \psi \psi
+ \textrm{h.c.}\right)
+ \dots\,,
\label{eq:L3}
\end{align}
\begin{marginnote}[]
\entry{NREFT Lagrangian}{Expansion in powers of $|\vec p \,|^2/m^2$ including LECs}
\end{marginnote}
where h.c. indicates hermitian conjugate,
$\overleftrightarrow{\nabla}= (\overrightarrow{\nabla}-\overleftarrow{\nabla})/2$
is the Galilean-invariant derivative,
and derivatives act only on adjacent objects.The field $\psi$ only destroys particles, since there is no corresponding antiparticle
in the theory.
NREFT is an expansion in $|\vec p \,|^2/m^2$, and thus
interaction terms are classified by the number of derivatives,
with the constraint from Bose symmetry that only an even number is allowed. Thus the
ellipses in Eqs.~(\ref{eq:L2}) and (\ref{eq:L3}) indicate terms with 
four or more derivatives, at which order arise the first terms that lead to $d$-wave interactions
between pairs of particles.
The effects of virtual particle-antiparticle pairs 
in the underlying relativistic theory
are subsumed into the {\em a priori}
unknown (dimensionful) constants $C_i$ and $D_i$.
We refer to these generically as LECs---low-energy constants. 

This theory has been extensively studied in infinite volume, e.g. in its application
to the three nucleon system (where the fields become fermionic and have a spin index).
For a review, see Ref.~\cite{BvK02}.
The LECs are to be determined by solving the two- and three-particle scattering and
bound-state problems 
(involving the solution of integral equations)
and comparing to experimental results for phase shifts and bound-state energies.
One must choose a regularization, a topic with an extensive literature given the
subtleties that arise in the presence of large scattering lengths~\cite{KSW,Phillips}.
For numerical applications, however, the standard choice is a hard cutoff, 
$|\vec p \,| < \Lambda$, and this is what is used in Refs.~\cite{Akaki1,Akaki2}.

The overall strategy employed in this approach is as follows.
The NREFT is to be solved in finite volume, working first with only the
leading order couplings $C_0$ and $D_0$, and including higher-order terms as
needed. 
These LECs are to be determined
by comparing the theoretically predicted spectrum to that obtained in a lattice calculation.
In a second step, the NREFT is then solved in infinite-volume, predicting 
the three-particle scattering amplitudes and bound-state properties.
This approach makes use of a crucial property of the LECs, namely
that they are expected to be volume-independent, since they arise from integrating
out short-distance physics. Thus the values obtained in finite volume can be applied
unchanged in the infinite-volume calculations.

This two-step strategy is similar to that used in the RFT analysis
presented earlier, where the intermediate, cutoff-dependent quantity $\Kdf$ was needed,
and infinite-volume quantities were obtained by solving integral equations.
In the NREFT approach the intermediate quantities are the LECs, which are also
cutoff dependent, and thus not directly physical.
What the NREFT approach provides in addition is a systematic power-counting
scheme, valid as long as one works in the NR regime.
\begin{marginnote}[]
\entry{Advantage of NREFT}{Systematic power-counting scheme}
\end{marginnote}

An important technical point discussed in Ref.~\cite{Akaki2} concerns
a class of higher-order terms in $\cL_2$ and $\cL_3$ that lead to vanishing 
on-shell contributions to physical scattering amplitudes at tree level.
An example from $\cL_2$ is a term leading to a vertex proportional to
$(\vec k^2-\vec p^2)^2$, where $\vec k$ and $\vec p$ are the relative momenta
between the two particles in initial and final states, respectively.
It is argued that one can choose the regularization such
that these terms do not contribute to physical quantities
even when included in loop diagrams, and thus that they can be dropped from the beginning.
In this way the intermediate quantities in the NREFT
approach provide a complete description of the physical amplitudes.
This is the analog here of the result in the RFT formalism that the intermediate
quantity $\Kdf$ is an on-shell amplitude. 

\begin{marginnote}[]
\entry{Technical result}{Only physical LECs need to be included}
\end{marginnote}
The restriction to ``physical'' LECs is implemented in Ref.~\cite{Akaki2}
using auxiliary dimer fields.
These are simply a technical device in which a composite field is introduced for
two of the particles for each choice of their relative angular momentum
(which here is constrained to be even).
Integrating out the infinite tower of
dimer fields leads back to the original Lagrangian, Eq.~(\ref{eq:L}),
but with only the physical LECs. 
The angular momentum of the dimer corresponds exactly to the indices $\ell$ and $m$
in the RFT approach.
Details of the implementation of the dimer fields
can be found in Ref.~\cite{Akaki2}.
\begin{marginnote}[]
\entry{Technical implementation}{Use of auxiliary dimer fields}
\end{marginnote}

The NREFT quantization condition has been worked out explicitly so far 
only for the case of an $s$-wave dimer.
This is the analog of truncating the RFT formalism to $\ell=0$,
an approximation described in Sec.~\ref{sec:trunc} above. 
This approximation requires that the two-particle scattering amplitude vanish for
all higher waves, and that the three-particle interaction does not couple to higher waves
in a pair. 
A further restriction introduced to simplify the derivation is that $\vec P=0$.

\begin{figure}
\includegraphics[width=\textwidth]{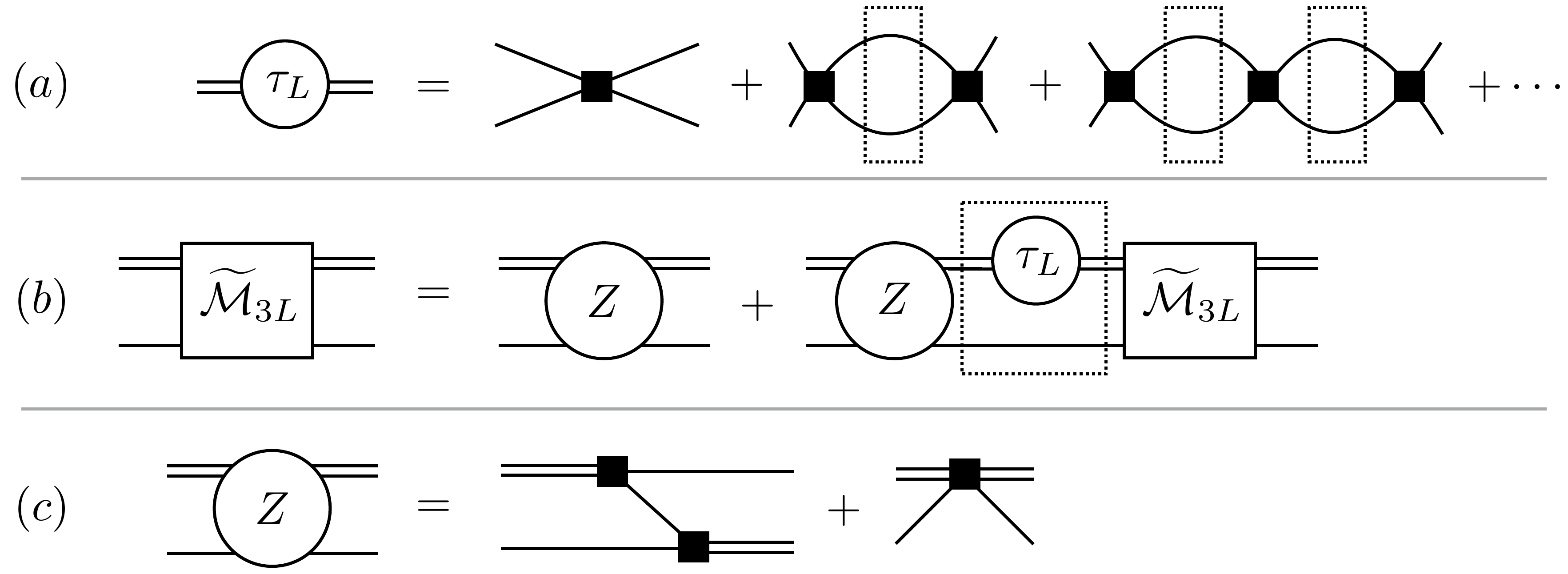}
\caption{Diagrammatic representation of components of NREFT derivation of the
quantization condition. As above, a dashed rectangle indicates that the
three-momentum in the loop is summed over finite-volume values.
Single lines are NREFT particle propagators, which are always forward in time.
Double lines indicate the dimer propagator.
(a) The finite-volume dimer propagator $\tau_L$, with 
filled squares representing the contributions from the interactions in $\cL_2$
[see Eq.~(\ref{eq:L2})].
(b) Integral equation for the dimer-particle scattering amplitude.
(c) The three-particle interaction kernel $Z$, with filled squares representing
contributions arising ultimately from interactions in $\cL_3$.}
\label{fig:NREFT}
\end{figure}

\subsubsection{Two-particle quantization condition}
The first step in the derivation of the three-particle quantization condition
is to determine the dimer propagator in finite volume, denoted
$\tau_L(\vec k)$ (with dependence on the total energy $E$ kept implicit). 
Here $\vec k$ is the spectator momentum, $\vec k$, 
which determines the dimer momentum to be $\vec P-\vec k=-\vec k$.
$\tau_L$ is proportional to the finite-volume scattering amplitude, $\cM_{2L}$, discussed in
Sec.~\ref{sec:QC2}. If we denote the $s$-wave component of the latter quantity,
evaluated in the NR regime,\footnote{%
We use the phrase ``in the NR regime'' here and below to indicate that we keep only
terms linear in $E_{\text{NR}}=E-3m$ and $\vec k^2/(2m)$.}
by $\cM_{2L,s}^\NR(P-k)$, where the argument denotes the dimer four-momentum,
then the precise relation is
\begin{equation}
32 \pi m \tau_L(\vec k) = \cM_{2L,s}^\NR(P-k)\,.
\label{eq:tauLtoM2L}
\end{equation}
The dimer propagator is given by the diagrams shown in {\bf Figure~\ref{fig:NREFT}(a)}.
Summing the geometric series leads to
\begin{equation}
\tau_L(\vec k)^{-1} = f_\tau(q_{k}^{*2}) + \frac{4\pi}{L^3} \sum_{\vec a}
\frac1{m E_{\rm NR} - \vec k^2 - \vec a^2 - \vec k \cdot \vec a}\,.
\label{eq:tauL}
\end{equation}
\begin{marginnote}[]
\entry{Finite-volume dimer propagator}{}
\end{marginnote}
Here $f_\tau$ arises from the vertices in $\cL_2$, and thus is a known
function of the $C_i$. It is proportional to the inverse of the BS kernel.
It depends on the squared relative c.m. momentum of the particles in the dimer,
\begin{equation}
q_{k}^{*2} = \tfrac14 \left[(E-\omega_k)^2 - \vec k^2 - 4 m^2\right]\,.
\label{eq:q2kstar}
\end{equation}
Finally, $E_{\rm NR}=E-3m$ is the nonrelativistic energy,
while $\vec a$ is summed over the allowed finite-volume values.

We see here how the NR limit simplifies the analysis compared to that
outlined in Sec.~\ref{sec:QC2}.
Here there are  only $s$-channel loops, and these contain only two particles.
All the other loops in RFT collapse to the point-like interactions parametrized by the LECs.
In particular, the BS kernel required in  Sec.~\ref{sec:QC2}
is replaced here by the function $1/[32\pi m f_\tau]$.
The nontrivial effort required to show that this kernel has only exponentially suppressed
volume dependence in the RFT approach is replaced here by the
assumed general result from EFT that LECs are independent of volume.\footnote{%
This discussion shows that the LECs are, in general, not strictly independent of
volume but instead can have an exponentially-suppressed dependence.}
With this result in hand, one simply determines the volume dependence implicitly by
calculating the dimer propagator with the loop integrals replaced by sums.

\begin{marginnote}[]
\entry{Advantage of NREFT}{Simplified analysis of finite-volume effects}
\end{marginnote}

The function $f_\tau$ can be determined by considering the dimer propagator in infinite
volume, $\tau(\vec k)$, obtained from Eq.~(\ref{eq:tauL}) by changing the sum
into an integral with an $i\epsilon$ prescription for the pole. 
Regulating the ultraviolet divergence in some manner,
the integral evaluates to
\begin{equation}
I_{\epsilon}(q_k^{*2}) \equiv
4\pi\,\int_{\vec a} \frac1{m E_{\text{NR}} - \vec k^2 - \vec a^2 - \vec k \cdot \vec a + i \epsilon}
= 4\pi \int_{\vec a^*} \frac1{\vec q_{k}^{*2} - a^{*2} + i \epsilon}
= \sqrt{-q_{k}^{*2}} + I_\PV(q_k^{*2})\,.
\label{eq:INR}
\end{equation}
Here $\int_{\vec a}\equiv \int d^3a/(2\pi)^3$,
and $\vec a^*$ is the result of boosting $\vec a$ to the dimer c.m. frame.
$I_\PV$ is the same integral except 
defined using the principal-value (PV) pole prescription.
Thus $I_\PV$ is a real, analytic function of $q_{k}^{*2}$,
which in fact evaluates to a constant.
The square-root in Eq.~(\ref{eq:INR}) 
is defined to have a negative imaginary part above threshold.
Using these results we obtain
\begin{equation}
\tau(\vec k)^{-1} = f_\tau(q_{k}^{*2}) + I_\PV + \sqrt{-q_{k}^{*2}}\,,
\end{equation}
which, when compared to the known form of the scattering amplitude 
[see Eq.~(\ref{eq:unitarity}) above]\footnote{%
The relation between the NR and relativistic versions of $\cK_{2,s}$
is purely kinematical, and given by
\begin{equation}
\cK_{2,s}(P-k) = \sqrt{1 + q_k^{*2}/m^2} \, \cK_{2,s}^\NR(q_k^*)\,.
\label{eq:K2NRvsK2}
\end{equation}}
\begin{equation}
\tau(\vec k)^{-1} = {32\pi m}\left[\cM_{2,s}^\NR(P-k)\right]^{-1}
= 32\pi m \left[\cK_{2,s}^\NR(q_{k}^*)\right]^{-1} + \sqrt{-q_{k}^{*2}}\,,
\end{equation}
leads to the conclusion that
\begin{equation}
f_\tau(q_{k}^{*2}) + I_\PV = 32\pi m\left[ \cK_{2,s}^\NR(q_{k}^*)\right]^{-1}\,.
\label{eq:ftaures}
\end{equation}
This shows explicitly how the LECs $C_i$, contained in $f_\tau$, are related
to the physical quantity $\cK_{2,s}$, albeit in a cutoff dependent manner.\footnote{%
In Ref.~\cite{Akaki2}, the integral is regulated by dimensional regularization, in
which case $I_\PV$ vanishes. However, since the sums are regulated in a different
manner (using a hard cutoff) we find it more consistent to use the same cutoff throughout,
and thus keep $I_\PV$ nonvanishing. This allows us to verify explicitly that the
final results of Ref.~\cite{Akaki2} are regulator independent.}

Combining Eqs.~(\ref{eq:tauL}) and (\ref{eq:ftaures}) we obtain the final result
for $\tau_L$,
\begin{align}
\left[{32\pi m} \tau_L(\vec k)\right]^{-1}
&= \cK_{2,s}^\NR(q_{k}^*)^{-1} + \frac1{8m}
\left[\frac1{L^3}\sum_{\vec a} - \PV\,\int_{\vec a}\right]
\frac{1}{m E_{\rm NR} - \vec k^2 - \vec a^2 -\vec k\cdot \vec a}\,,
\\
&= \cK_{2,s}^\NR(q_{k}^*)^{-1} + F_{2,s}^\NR(P-k,L)\,.
\label{eq:tauLfinal}
\end{align}
The first equality is our preferred way of writing Eq.~(3.2) of Ref.~\cite{Akaki2},
as it shows that the volume-dependence of $\tau_L$ arises from a sum-integral 
difference, just as in the analysis in Sec.~\ref{sec:QC2}.\footnote{%
The result quoted in Ref.~\cite{Akaki2} contains only the sum, 
but is equivalent to that here since the corresponding integral vanishes in the
regularization used in that work.}
To obtain the second equality, we note that the sum-integral difference
is simply the $s$-wave component of $F_2(P-k,L)$, Eq.~(\ref{eq:Fdef}), 
evaluated in the NR regime,
and with the $i\epsilon$ pole prescription replaced by the PV prescription.
Thus we call it $F_{2,s}^\NR$, which we emphasize is a real quantity.

The two-particle quantization condition can now be obtained
as an intermediate result.
Energy levels are given by the positions of poles in $\cM_{2L}$, 
leading to the algebraic result
\begin{equation}
\tau_L^{-1} = 0 
\ \ \Rightarrow \ \
(\cK_{2,s}^\NR)^{-1} + F_{2,s}^\NR = 0\,.
\end{equation}
This is equivalent to the result derived in Sec.~\ref{sec:QC2},
 Eq.~(\ref{eq:QC2K}), when one keeps only the $s$-wave component 
 and works in the NR regime.
 \begin{marginnote}[]
\entry{Two-particle quantization condition in NREFT}{}
\end{marginnote}

\subsubsection{Three-particle quantization condition}
This is derived in Ref.~\cite{Akaki2} by considering a quantity,
$\widetilde \cM_{3L}$,  the particle-dimer scattering amplitude.
This is closely related to the $s$-wave restriction of the
finite-volume three-particle amplitude, $\cM_{3L}$,
introduced in Sec.~\ref{sec:KtoM}.\footnote{%
Strictly speaking, $\widetilde \cM_{3L}$ is most closely related to the
unsymmetrized amplitude $\cM_{3L}^{(u,u)}$ defined in Refs.~\cite{HS1,HS2}.}
To obtain $\cM_{3L}$ from
$\widetilde \cM_{3L}$, one adds vertices at each end connecting the dimer to
two particles, and then symmetrizes.
This implies that, as for $\cM_{3L}$, the poles of $\widetilde \cM_{3L}$
occur at the energies of three-particle finite-volume states, so that it is a
good quantity to consider to derive the quantization condition.
In NREFT, $\widetilde \cM_{3L}$ satisfies 
\begin{equation}
\widetilde\cM_{3L;pk} = Z_{pk} + 
\frac{8\pi}{L^3} \sum_{\vec q} Z_{pq}\tau_L(\vec q) \widetilde\cM_{3L;qk}\,,
\label{eq:M3Lt}
\end{equation}
\begin{marginnote}[]
\entry{Matrix equation leading to three-particle quantization condition in NREFT}{}
\end{marginnote}
which is shown schematically in  {\bf Figure~\ref{fig:NREFT}(b)}.
Here, the subscripts $p$, $q$ and $k$ are shorthands for the 
corresponding spectator momenta,
while the kernel $Z_{pq}$, shown in {\bf Figure~\ref{fig:NREFT}(c)},  is
\begin{equation}
Z_{pq} = Z^0_{pq} + \frac{H_0(\Lambda)}{\Lambda^2} + \dots \,,
\qquad
Z^0_{pq} =
\frac1{\vec p^{\,2} + \vec q^{\,2} +\vec p \cdot \vec q - m E_{\text{NR}}} \,.
\label{eq:Z}
\end{equation}
Here $H_0$ is a dimensionless constant proportional to $D_0/C_0^2$, 
$\Lambda$ is the hard cutoff on the sums,
and the ellipsis represents contributions from higher order terms
in $\cL_3$, proportional to $D_2$, etc.

To use Eq.~(\ref{eq:M3Lt}) one assumes that $\cK_{2,s}$ has been determined 
using the two-particle quantization condition, and thus that $\tau_L$ is known.
Since the sum over $\vec q$ is cut off, Eq.~(\ref{eq:M3Lt}) is
a finite matrix equation  for $\widetilde \cM_{3L}$, which can be solved numerically
for a given choice of the LECs. One then adjusts the LECs until the finite-volume
spectrum determined from a calculation in the underlying theory (i.e.~lattice QCD
if considering the $3\pi^+$ system) matches that given by Eq.~(\ref{eq:M3Lt}).
In practice, one should project onto irreps (irreducible representations) of the
symmetry group of the cubic lattice, as described in Ref.~\cite{DHMPRW},
which allows a more explicit formula for the quantization condition to be given.
We will, instead, provide an alternative explicit formula in the following, one that
shows more clearly the relationship to the RFT result derived in the previous section.

Once one has determined the LECs as just described, a second step is required
to predict the infinite-volume scattering amplitude $\cM_3$ and, from this, 
the properties of any bound states and three-particle resonances.
In the NREFT approach this step is simple: one uses Eq.~(\ref{eq:M3Lt})
but with $\tau_L$ replaced by $\tau$ and the sum replaced by an integral with an
$i\epsilon$ pole-prescription. This leads to the standard NREFT integral equation
for three-particle scattering, reviewed, for example, in Ref.~\cite{BvK02}.
This step is the analog of the integral equation relating $\Kdf$ to $\cM_3$
described in Sec.~\ref{sec:KtoM}.

We now comment briefly on the derivation of Eq.~(\ref{eq:M3Lt}).
This is conceptually just as straightforward as for two particles.
This is because, in NREFT,  all loops in a three-particle amplitude contain three particles, 
in constrast to a generic QFT in which (with a $Z_2$ symmetry) one can have five, seven,
etc. This allows one to calculate all the diagrams explicitly, without introducing
auxiliary objects such as the BS kernels.
In finite volume one simply replaces the integrals in these loops
with momentum sums (after the time-component integral is done).
LECs are again volume-independent, and summing all diagrams leads to
Eq.~(\ref{eq:M3Lt}).

\subsubsection{Relation to RFT approach}\label{sec:NREFTtoRFT}
It has been shown in Ref.~\cite{Akaki2} that the NREFT quantization condition
described implicitly above, and the RFT quantization condition of Eq.~(\ref{eq:QC}),
are algebraically equivalent when only the $s$-wave dimer is included in the latter and
if only the leading order, momentum-independent, two- and three-particle interaction
terms are kept, i.e.~if the only nonvanishing LECs are $C_0$ and $D_0$.
Here we give an alternative derivation of this result that is more direct and explicit.

We begin by noting that $Z^0$ is simply related to the switch factor $G$
given in Eq.~(\ref{eq:G}).
Denoting the $s$-wave ($\ell'=\ell=0$) part of $G$ in the NR regime
by $G_s^\NR$, one can easily show that
\begin{equation}
Z^0_{pq} = - 4 m L^3G_{s,pk}^\NR\,.
\label{eq:GtoZ}
\end{equation}
Here we have used the fact that the cutoff functions $H(\vec k)$ are unity to all orders
in the NR expansion.
To simplify subsequent manipulations we introduce the definitions
\begin{equation}
\overline \cM_{3L} \equiv \frac{\widetilde \cM_{3L}}{4 m L^3}\ \ {\rm and} \ \ 
\overline H \equiv \frac1{4 m L^3} \left(\frac{H_0(\Lambda)}{\Lambda^2} + \dots \right)
= \overline H_0 + \dots \,.
\label{eq:Hbar}
\end{equation}
In terms of these quantities Eq.~(\ref{eq:M3Lt}) can be rewritten as
\begin{equation}
\overline \cM_{3L} = ( -G_s^\NR + \overline H) + (-G_s^\NR+\overline H) \cM_{2L,s}^\NR
\overline \cM_{3L}\,,
\label{eq:M3Lbar}
\end{equation}
where we have also used Eq.~(\ref{eq:tauLtoM2L}).
This is a matrix equation, in which all spectator-momentum indices are implicit.
Solving, we find
\begin{align}
\overline \cM_{3L} &= \frac 1{1 - (-G_s^\NR + \overline H) \cM_{2L,s}^\NR} 
(-G_s^\NR + \overline H)\,,
\end{align}
which has a pole whenever
\begin{equation}
\det\left[1 - (-G_s^\NR+\overline H) \cM_{2L,s}^\NR\right] = 0 \,.
\end{equation}
This can be rewritten using Eqs.~(\ref{eq:tauLtoM2L})
and (\ref{eq:tauLfinal}) as
\begin{equation}
\det\left[(\cK_{2,s}^\NR)^{-1} + F_{s}^\NR + G_s^\NR -\overline H \right] = 0 \,.
\label{eq:QC3NR}
\end{equation}
\begin{marginnote}[]
\entry{Three-particle quantization condition in NREFT}{}
\end{marginnote}
Here, as in Sec.~\ref{sec:QC3}, we have elevated $\cK_{2,s}^\NR$ 
into a diagonal matrix with entries $\cK_{2,s}^\NR(q_k^*)$, 
while, following Eq.~(\ref{eq:F}), we denote the matrix form of $F_{2,s}^\NR$
by $F_{s}^\NR$.
Equation~(\ref{eq:QC3NR})
is the NREFT quantization condition of Refs.~\cite{Akaki1,Akaki2} expressed
in notation similar to that of Ref.~\cite{HS1}.
Note that the expansion of $\cK_{2,s}$ and $\overline H$ in powers of momentum
has not been truncated at this stage, i.e. all the $C_i$ and $D_i$ are still included.
The determinant is finite-dimensional because of the hard cutoff applied to the
implicit spectator-momentum indices.

We now compare this to the quantization condition of Ref.~\cite{HS1},
given in Eq.~(\ref{eq:QC}).
In the $s$-wave approximation, this has the form
\begin{equation}
\det\left[F_{3,s}^{-1} + \Kdfs\right] = 0\,,
\quad
F_{3,s} = \frac{F_{s}}{2\omega L^3} \left[
\frac13 - \frac1{\cK_{2,s}^{-1} + F_{s} + G_s} F_{2,s}\right]\,.
\label{eq:F3s}
\end{equation}
Here the subscript $s$ indicates keeping only $\ell'=\ell=0$ contributions,
so all quantities are matrices with only spectator-momentum indices.
Note that, at this stage, the NR limit has not been taken, so the sums are cut off
by the functions $H(\vec k)$ in $F_{s}$ [see Eq.~(\ref{eq:F})] and $G_s$ [see Eq.~(\ref{eq:G})].
By straightforward algebraic manipulations, 
Eq.~(\ref{eq:F3s}) can be written in a form that looks similar to the NREFT
result, Eq.~(\ref{eq:QC3NR}):
\begin{equation}
\det\left[\cK_{2,s}^{-1} + F_{s} + G_s - \overline H^R\right]=0\,, \quad
\overline H^R = -(\cK_{2,s}^{-1} + G_s - 2F_{s}) \Kdfs \frac{F_{s}}{6\omega L^3} \,.
\label{eq:HSQC2}
\end{equation}
\begin{marginnote}[]
\entry{Relativistic quantization condition}{restricted to $s$-wave}
\end{marginnote}
Thus for the two quantization conditions to agree three conditions must be satisfied: (i) we must consider 
Eq.~(\ref{eq:HSQC2}) in the NR regime (so that $G_s\to G_s^\NR$, etc.), (ii) we must use a hard cutoff in this equation, {\em and} (iii) we must demonstrate $\overline H = \overline H^R$.

We consider these requirements for equivalence in turn.
The spectator momenta in Eq.~(\ref{eq:HSQC2}) run up to
a smooth (rather than hard) cutoff at $\Lambda^R\sim m$.
In principle, one could reduce $\Lambda^R$ into the NR regime so that
$G_{s} \to G_{s}^\NR$ etc.
Then, setting aside the issue of $\overline H$ vs. $\overline H^R$,  
the only difference between the results
would be that between a hard and a smooth cutoff.

In practice, however, reducing $\Lambda^R$ into the NR regime is problematic.
Present calculations using LQCD are done with $m_\pi L\sim 4-6$ 
and $m_\pi \approx m_\pi^{\rm phys}$.
With these parameters, even the first excited state lies outside
the NR regime:
\begin{equation}
{E_1}/{m_\pi} = \sqrt{1 + [{2\pi}/({m_\pi L})]^2} \sim 1.5-1.9
\,.
\label{eq:relkin}
\end{equation}
Thus a practical quantization condition should have its cutoff at a relativistic energy
and include relativistic kinematics.
In this regard, we note that Ref.~\cite{Akaki2} argue that the NREFT quantization
condition can be ``relativized'' by including the correct kinematical factors.
Indeed, we see that this can be accomplished in the present instance by
replacing each of the NR quantities by their relativistic counterparts, as introduced in the RFT approach.
\begin{marginnote}[]
\entry{Need for relativistic kinematics in practice}{}
\end{marginnote}

The final requirement for the agreement between the two quantization conditions,
$\overline H=\overline H^R$,
is, at first sight, more problematic. $L^3 \overline H$ is an infinite-volume quantity,
while $L^3\overline H^R$ is not, since it contains $G_s$ and $F_{s}$. In addition,
$\overline H_{pq}$ is, by construction, a smooth function of $\vec p$ and $\vec q$,
while the presence of $G_s$ and $F_{s}$ implies that there are singularities in
$\overline H^R$.\footnote{%
Of course, these quantities are evaluated only for discrete, finite-volume momenta,
so one will in general not hit the singularities, but the point here is that the two
quantities appear to have very different momentum dependence.}
It turns out that these issues can be resolved if one (a) takes the NR limit
of Eq.~(\ref{eq:HSQC2}), (b) assumes that $\cK_{2,s}^\NR$ is independent of momentum,
which is equivalent to keeping only the leading $C_0$ term in 
the NR expansion of $\cL_2$, i.e. keeping
only the scattering length in the effective range approximation, and (c) 
assumes that $\Kdfs$ and $\overline H$ are
independent of spectator momenta.
For $\Kdfs$, this is the isotropic limit that holds near threshold, as
discussed in Sec.~\ref{sec:trunc}.
For $\overline H$ this means keeping only the leading $D_0$ term in
the NR expansion of $\cL_3$.
In other words, we consistently keep only the leading-order terms in the NR limit.

In this combined limit, $\Kdfs = | 1\rangle \Kiso \langle 1 |$, 
with $|1\rangle$ the unnormalized
isotropic vector having a unit entry in all positions.
This allows us to use the following identity
\begin{equation}
G_s^\NR |1\rangle = 2 F_{s}^\NR |1\rangle + I_s^\NR  |1\rangle \,,
\ \ {\rm with} \ \
I^\NR_{s;pk} = \delta_{pk} \frac{I_\PV}{16 \pi m}\,,
\label{eq:GFid}
\end{equation}
\begin{marginnote}
\entry{Key identity between $F$ and $G$}{Valid in isotropic approximation}
\end{marginnote}
together with its transpose. Recall that $I_\PV$ is a momentum-independent,
regularization-dependent constant.
Denoting the leading contribution to $\overline H^R$ in the NR limit by
$\overline H^\NR$, we then find
\begin{equation}
\overline H^\NR = -
([\cK_{2,s}^\NR]^{-1} + I_s^\NR) \Kdfs
\frac{[\cK_{2,s}^\NR]^{-1}+F_{s}^\NR+G_s^\NR - I_s^\NR -[\cK_{2,s}^\NR]^{-1}}{18m L^3}
\,.
\end{equation}
Inserting this result into the NR form of Eq~(\ref{eq:HSQC2}), 
the matrix inside the determinant can be written
\begin{multline}
\left[1 + \frac{(\cK_{2,s}^\NR)^{-1} + I_s^\NR}{18 m L^3} \Kdfs \right]
\left[(\cK_{2,s}^\NR)^{-1} + F_{s}^\NR + G_s^\NR\right]  
-\\  \frac1{18 mL^3}\left[(\cK_{2,s}^\NR)^{-1} + I_s^\NR\right] \Kdfs 
\left[(\cK_{2,s}^\NR)^{-1} + I_s^\NR\right] \,.
\end{multline}
This allows the leading NR term in the relativistic quantization condition to be written
in exactly the form of the NR quantization condition, Eq.~(\ref{eq:QC3NR}), with
\begin{equation}
\overline H =  \overline H_0 = 
\left[1 + \frac{(\cK_{2,s}^\NR)^{-1} + I_s^\NR}{18 m L^3} \Kdfs \right]^{-1}
[(\cK_{2,s}^\NR)^{-1} + I_s^\NR] \frac{\Kdfs}{18 mL^3} [(\cK_{2,s}^\NR)^{-1} + I_s^\NR]\,.
\label{eq:Hbarres}
\end{equation}
\begin{marginnote}
\entry{Relation of RFT and NREFT quantization conditions}{$s$-wave dimers only, and in the NR limit}
\end{marginnote}
Using the fact that $\cK_{2,s}^\NR$ and $I_s^\NR$ are proportional to the
identity, it is simple to show that this result is isotropic, as required for complete 
equivalence. The overall factor of $1/(mL^3)$ also matches that in $\overline H_0$
[see Eq.~(\ref{eq:Hbar})].
Furthermore, since for any constant $x$, we have
\begin{equation}
\frac1{1 + x |1\rangle \langle 1 |} |1\rangle = |1\rangle \frac1{1 +x N}\,,
\quad
N = \langle 1 | 1\rangle \propto (L\Lambda)^3\,,
\end{equation}
we see that the $L^3$ in the first factor in Eq.~(\ref{eq:Hbarres}) cancels,
so that the right-hand side is a volume-independent constant, as required to
match $\overline H_0$.\footnote{%
More precisely, this is true up to corrections suppressed by powers of $1/(\Lambda L)$.
}
The presence of the cutoff dependent quantity $I^\NR$ on the right-hand side is not
an issue, because both $H_0$ and $\Kdfs$ are cutoff dependent.
Indeed, this result allows one to map the cutoff dependence of $H_0$ known from NREFT
to that of $\Kdfs$.

\subsubsection{Summary} The derivation of the three-particle quantization
condition is dramatically simplified by using NREFT, compared to the RFT derivation
described earlier. This is partly due to the fact that, so far, only the $s$-wave dimer has
been included in the former. But the primary reasons for the simplicity are
(a) that the number of diagrams is sufficiently small
that one can straightforwardly include them all in a simple and explicit integral equation;
and (b) that the two-step approach using intermediate quantities is embraced from
the beginning. No attempt is made to explicitly find
all sources of power-law volume dependence by focusing on sum-integral differences.
Instead, one simply uses the same NREFT in separate finite- and infinite-volume
calculations. Given point (a) above, this can be done without further approximation. 
An additional advantage of the NREFT approach is that it is valid also in the
presence of subchannel resonances, since any volume-dependence they introduce
is included automatically.

The simplicity of the NREFT approach is not available in a generic RFT, since one cannot
solve the three-particle scattering problem in a generic theory. 
Thus one is forced to keep track
of finite-volume effects explicitly, leading to a more complicated derivation, and 
with additional considerations required for subchannel resonances.

Of course, if both approaches are carried out correctly, they should agree
when we take the NR limit of the relativistic approach. This is indeed the case,
as first discussed in Ref.~\cite{Akaki2}, and as shown explicitly earlier in this section.

The main drawback with the NREFT approach is simply that most three-body systems
of interest in nuclear and particle physics are relativistic. We have already commented
on the kinematics of three pions [see Eq.~(\ref{eq:relkin})]: for the volumes used
in lattice QCD calculations the sum over spectator momenta necessarily lies
in the relativistic domain. This conclusion appears unavoidable for
applications of the three-particle quantization condition to lattice QCD.

A further comment on the NREFT approach is that, once interaction terms 
quartic in derivatives are included, i.e.~once $C_4$ and $D_4$ are nonzero,
then one must also include $d$-wave ($\ell=2$) dimers, as they enter at the
same power in the NR expansion. 
As one goes further into the relativistic domain, 
many higher-order terms will be needed, and thus
many higher-order dimers must be included.

In summary, for the NREFT approach to have broad utility, it is necessary both
that the formalism be explicitly extended to include higher waves (which we expect
to be relatively straightforward) and that the kinematics somehow be relativized.
This second step is claimed also to be straightforward in Ref.~\cite{Akaki2}.
We think, however, that it will be important at each stage to check that the results
agree with those from the RFT approach.

\subsection{Finite-volume unitarity (FVU) approach}
\label{sec:FVU}
The third approach that has been used to derive a three-particle quantization condition
aims to maintain relativistic invariance but to avoid 
much of the work of the diagrammatic RFT approach by using the
constraints arising from unitarity~\cite{MD1}.
The starting point is a representation of $\cM_3$ in terms of dimers\footnote{%
In Refs.~\cite{MD0,MD1} the dimers are referred to as ``isobars'', but they play essentially
the same technical role in the analysis, so we prefer to use the name dimer throughout.}
that is explicitly unitary in the $s$-channel~\cite{MD0}.
Then, by a judicious replacement of integrals with sums, a representation of
a quantity similar to $\cM_{3L}$ is obtained, and from this follows the quantization condition.
As in the NREFT approach,
the formalism has been worked out so far only for $s$-wave dimers,
and for $\vec P=0$.

\subsubsection{Two-particle quantization condition}
\label{sec:FVU2}
The discussion begins by considering the two-particle subsystems.
The infinite-volume $s$-wave scattering amplitude is written as~\cite{MD0}
\begin{equation}
\cM_{2,s} \equiv - v(q_1,q_2) \frac1{D(s_{12})} v(p_1,p_2)\,,
\label{eq:MDM2s}
\end{equation}
where $p_i$ and $q_i$ are the initial and final momenta, respectively, of
the scattering pair, and $s_{12}=(p_1+p_2)^2$.
The $s$-channel cut lies in $D$, which is thus complex, while $v$ is a smooth, real
function. By relativistic invariance, it can depend only on $s_{12}$, but it is convenient
to write it as a function of $Q^2 = -(p_1-p_2)^2 = s_{12} - 4 m^2$ (using the mostly-minus
metric).
The form used in Ref.~\cite{MD2} is
\begin{equation}
 v(p_1,p_2) = \lambda(s_{12}) f(Q^2)\,,\quad
 f(Q^2) = \frac{1}{1 + \exp\left[Q^2/4-(1-\Lambda/2)^2\right]}\,,
 \label{eq:v}
 \end{equation}
where $\lambda$ is a smooth function and $\Lambda$ a parameter. 
Unitarity determines the imaginary part of the denominator $D$, and the full quantity
can be reconstructed using an appropriately subtracted
dispersion relation. The result can be written\footnote{%
This is equivalent to the form given in Eq.~(3) of Ref.~\cite{MD2}
after correcting a typographical error.}
 \begin{equation}
 D(s_{12}) = s_{12} - M_0^2
 - \frac{\lambda(s_{12})^2}2
  \int_{\vec q} \frac1{2\omega_q}
 \frac{f(4 \vec q^{\,2})^2}{s_{12} - 4\omega_q^2+ i \epsilon}
 \,,
 \label{eq:D}
 \end{equation}
with $M_0$ an additional parameter.
Thus we see that the introduction of the form factor in Eq.~(\ref{eq:MDM2s})
leads to its appearance as a convergence factor in the loop integral in $D$.
The claim is that this form for $D$, when inserted into Eq.~(\ref{eq:MDM2s}),
gives the most general unitary result for $\cM_{2,s}$.
In applications, the form for $\lambda(s_{12})$ must be tuned so as to match 
the known phase shift.

The next step is to claim that the finite-volume amplitude $\cM_{2L}$ is obtained
simply by replacing the integral in Eq.~(\ref{eq:D}) with the finite-volume momentum sum.
Although the integral is frame-invariant, the sum depends on the choice of frame.
Anticipating the three-particle application, we label the frame
by the spectator momentum, $\vec k$
(so that the dimer momentum is $-\vec k$):
 \begin{align}
 \cM_{2L,s}(\vec k) &= - v_\on(\vec k) \frac1{D_L(\vec k)} v_{\on}(\vec k)\,,
 \label{eq:MDM2L}
 \end{align}
 Here $v_\on$ is the same vertex function as appearing in Eq.~(\ref{eq:MDM2L}), 
but the new subscript is used to emphasize that external particles are on shell.
The expression for $D_L(\vec k)$ for general $\vec k$ is given in Ref.~\cite{MD1},
but here we show only the form for  the dimer at rest:
 \begin{align}
 D_L(\vec k=0)  &= E_{2,k}^{*2} - M_0^2
 - \frac{\lambda(E_{2,k}^{*2})^2}2
 \frac1{L^3}\sum_{\vec q} \frac1{2\omega_q}
 \frac{f(4 \vec q^{\,2})^2}{E_{2,k}^{*2} - 4\omega_q^2+ i \epsilon}\,,
 \label{eq:DL}
\end{align}
where $E_{2,k}^{*2} = (E-\omega_k)^2-\vec k^2$ is the value of $s_{12}$.
The two-particle quantization condition for a general frame is then obtained
from the poles of $\cM_{2L}$. Since the vertex function is nonsingular,
the poles occur when
\begin{equation}
D_L(\vec k) = 0\,.
\label{eq:MDQC2}
\end{equation}
This is the $s$-wave two-particle quantization condition, which 
can be used to constrain the parameters in $\lambda(s)$ and $f(Q^2)$
given the finite-volume spectrum.
\begin{marginnote}
\entry{FVU approach}{2-particle quantization conditions}
\end{marginnote}

This approach can be justified by the following argument, which we describe in
some detail as we have not found it given explicitly in the literature.
As discussed in Sec.~\ref{sec:QC2}, power-law finite-volume behavior results
only from sum-integral differences over singular summands/integrands.
Unitary cuts pick out exactly those loops for which integrands have poles,
because it is only by integrating across a pole (with an $i\epsilon$ prescription) that
one can obtain an imaginary part.
In the present case, this can occur only for two-particle loops (as long as $s_{12} < 16 m^2$).
Thus the unitary cuts  pick out exactly those loops whose sum-integral difference
leads to power-law volume effects. By writing $\cM_2$ in the form
given by Eqs.~(\ref{eq:MDM2s}) and (\ref{eq:D}) one is able to isolate the contributions
from such loops. The infinite-volume quantities in these expressions,
 $M_0$, $v$, $\lambda(s)$ and $f(Q^2)$, certainly involve loop contributions,
 but these do not have singular integrands, and so sum-integral differences are
 exponentially suppressed. Thus they can be used unchanged in the
expression for $\cM_{2L}$, Eq.~(\ref{eq:DL}).

We find this argument very plausible, but to be completely convinced of the conclusion
we rely on the fact that the quantization condition in Eq.~(\ref{eq:MDQC2}) 
can be shown to be equivalent to that derived above in the RFT approach.
This is straightforward to show, as noted in Ref.~\cite{MD2},
but it is useful to provide the explicit argument as the result will be used later.
The key point is that $D_L$ and $D$ differ only
by a sum-integral difference, and this can be rewritten in terms of 
the L\"uscher zeta-function $F_{2,s}^\ieps$. 
We find that, for any choice of $\vec k$,
\begin{equation}
D_L(\vec k) = D(\vec k) -  v_{\rm on}(\vec k) F^{i\epsilon}_{2,s}(\vec k)  v_{\rm on}(\vec k)\,.
\label{eq:DLvsD}
\end{equation}
Note that the sum-integral difference projects
the function $f$ to its on-shell value, leading to $F_{2,s}^\ieps$ being sandwiched
between on-shell vertex functions. 
Using Eq.~(\ref{eq:MDM2s}), we can rewrite $D_L$ as
\begin{equation}
D_L(\vec k) = -  v_{\rm on}(\vec k) 
\left[\cM_{2,s}(\vec k)^{-1} + F^{i\epsilon}_{2,s}(\vec k) \right] v_{\rm on}(\vec k)\,.
\label{eq:DLa}
\end{equation}
Finally, since $v_{\text{on}}$ is nonsingular, the quantization condition
(\ref{eq:MDQC2}) is equivalent to
\begin{equation}
0 = \cM_{2,s}(\vec k)^{-1} + F^{i\epsilon}_{2,s}(\vec k) = \cK_{2,s}(\vec k)^{-1}
+ F_{2,s}(\vec k)\,,
\label{eq:HSQCs}
\end{equation}
where we recall that $F_{2,s}$ differs from $F_{2,s}^\ieps$ by using the PV
pole prescription.
These results are identical to
the $s$-wave projections of the conditions given in Eqs.~(\ref{eq:QC2}) and
(\ref{eq:QC2K}).
We stress that  there are no caveats to this equivalence---it is an identity
that holds for all $\vec k$. 
\begin{marginnote}
\entry{Equivalence of FVU and RFT 2-particle quantization conditions}{}
\end{marginnote}

\subsubsection{Three-particle quantization condition}
In Ref.~\cite{MD0} a representation of the three-particle scattering amplitude, $\cM_3$,
is given in the presence of a single dimer. A key component is the particle-dimer scattering
amplitude, denoted $T(\vec p, \vec k)$, with $\vec p$ and $\vec k$ spectator momenta. 
This is the relativistic version of the quantity $\widetilde \cM_{3}$ 
appearing in the NREFT derivation.
It is shown in Ref.~\cite{MD0} that, if one assumes that $T$ satisfies a 
Bethe-Salpeter-type equation
\begin{equation}
T(\vec p,\vec k) = B(\vec p,\vec k) - \int_{\vec q}
B(\vec p,\vec q) \frac1{2\omega_q D(\vec q)} T(\vec q,\vec k)\,,
\label{eq:T}
\end{equation}
with $D$ the dimer propagator introduced above, then $\cM_3$ is unitary as long as $B$
takes the form
 \begin{equation}
 B(\vec p, \vec q) = B_0(\vec p, \vec q) + C(\vec p, \vec q)\,,\quad
B_0(\vec p, \vec q) \equiv  - 
 \frac{\lambda(E_{2,p}^{*2}) f([P-q-2p]^2)f([P-2q-p]^2)\lambda(E_{2,q}^{*2})}
 {(P-p-q)^2 -m^2 + i\epsilon}\,,
 \label{eq:B}
 \end{equation}
with $C$ a smooth function.\footnote{%
It is possible to change the definition of $B_0$ away from the pole, leading
to changes in the definition of $C$. Here we show the form from Ref.~\cite{MD0}.
Somewhat different choices are made in Refs.~\cite{MD1,MD2}.}
Comparing to the NREFT derivation, we observe that Eq.~(\ref{eq:T}) is the relativistic
analog of (the infinite-volume version of) Eq.~(\ref{eq:M3Lt}),
while   $B_0$ is the relativistic version of the kernel $Z^0$ defined in Eq.~(\ref{eq:Z}).
We stress that the one-particle exchange (OPE) form of $B_0$ follows from
enforcing unitarity, and not from calculating Feynman diagrams. There are no constraints
on $C$, other than smoothness.

An important question is whether this construction of $T$ is completely general.
In other words, while it has been shown that it results in a unitary $\cM_3$,
is the freedom left in the function $C$ sufficient to describe an arbitrary theory?
This question is not addressed in Ref.~\cite{MD0}. 
We shall assume in the following that the construction is general.

The next step in the derivation is to assert that, in order to obtain the finite-volume version
of $T$, and thus of $\cM_3$,
it is sufficient 
(up to exponentially-suppressed contributions)
to replace $D$ with $D_L$ and the integral
in Eq.~(\ref{eq:T}) with a finite-volume sum. 
Assuming so, then 
\begin{equation}
T_{L; p k} = B_{pk} - \frac1{L^3}\sum_{\vec q}
B_{pq} \frac1{2\omega_q D_L(\vec q)} T_{L;q k}\,,
\label{eq:TL}
\end{equation}
where we have written the quantities in matrix form, since the spectator momenta
are now discrete, e.g., $B_{pk} = B(\vec p, \vec k)$.
We stress again that there is an implicit dependence of all quantities on the overall
energy $E$.
In addition, the sum over $\vec q$ has to be cut off, and in Refs.~\cite{MD1,MD2} this
is done with a hard cutoff $|\vec q\,| < \Lambda$.
Then Eq.~(\ref{eq:TL}) is a matrix equation for $T_L$ that can be inverted.
Poles in $\cM_{3L}$ occur when $T_L$ has a divergent eigenvalue, 
which leads to the quantization condition
\begin{equation}
\det(T_L^{-1}) = 0\,.
\label{eq:MDQC3}
\end{equation}
This can be further reduced by projecting onto irreps of the cubic
 group~\cite{MD1,MD2,DHMPRW}, but the unreduced form will be sufficient here.

The derivation is structurally similar to that given in the NREFT approach, with the
LECs being replaced here by the unknown functions $C$, $\lambda$ and $f$.
What differs is that, whereas in the NREFT approach loops involve only three
particles, here, in any given relativistic theory,
there are loops involving any (odd) number of particles. The claim is that the
decomposition used above picks out all the three-particle loops that contain poles, 
because it is these loops that lead to the imaginary parts needed to satisfy unitarity.
These loops must be summed when in finite volume, while the 
other loops (contained inside $C$, $\lambda$ etc.) can be kept in infinite volume.
As for the two-particle quantization condition, this argument is very
plausible, but does not constitute in our view a complete derivation.
Thus we think that it is important to demonstrate the equivalence
of Eq.~(\ref{eq:MDQC3}) to the results of the RFT approach in the limit of including
only the $s$-wave dimer, as we do in the following.

\subsubsection{Relation to RFT approach}\label{sec:FVUtoRFT}
We begin by rewriting the FVU quantization condition. Defining $\omega$ and
$D_L$ as diagonal matrices as in the RFT derivation, we can solve Eq.~(\ref{eq:TL})
to find
\begin{equation}
T_L = \frac1{1+ B \frac{1}{2\omega L^3 D_L}} B
= 2\omega L^3 D_L \frac1{B + 2\omega L^3 D_L} B
\,.
\end{equation}
The quantization condition (\ref{eq:MDQC3}) can thus be rewritten as
\begin{equation}
\det \left( B + 2\omega L^3 D_L\right) = 0\,.
\label{eq:MDQC3a}
\end{equation}
We next note that 
\begin{equation}
B = - v_\on G_s v_\on (2\omega L^3) + C'\,,
\quad
C' - C = v_\on G_s v_\on (2\omega L^3) - B_0 \,,
\label{eq:Bnew}
\end{equation}
with $G_s$ the $s$-wave  part of $G$, Eq.~(\ref{eq:G}).
The key point here  is that the two terms in $C'-C$ 
have the same residue at the OPE pole, so that the difference cancels the pole
and is smooth.\footnote{%
Away from the pole, the two terms on the right-hand side of $C'-C$ differ
both because $G_s$ contains cutoff functions $H(\vec k)$ while $B_0$ does not,
but also because the vertices in $B_0$ are evaluated off shell.}
We also need the result from Eqs.~(\ref{eq:DLa}) and (\ref{eq:HSQCs}) that
\begin{equation}
D_L = - v_\on \left(\cK_{2,s}^{-1} + F_{s} \right)  v_\on
\,,
\label{eq:DLb}
\end{equation}
where $F_s$ is the matrix form of $F_{2s}$ [see Eq.~(\ref{eq:F})].
Putting this all together, and using the smoothness of $v_\on$,
the FVU quantization condition becomes
\begin{equation}
\det\left[\cK_{2,s}^{-1} + F_{s} + G_s
- \widetilde C (2\omega L^3)^{-1}  \right] = 0\,,
\quad
\widetilde C = v_\on^{-1} C' v_\on^{-1}\,.
\label{eq:MDQC3b}
\end{equation}
Aside from one technical point, to be discussed shortly,
this is an exact rewriting of the result of Ref.~\cite{MD1}.
This new result looks very similar to the form of the NREFT quantization condition
given in Eq.~(\ref{eq:QC3NR}),
with $\widetilde C$ here playing the role of $2 m L^3 \overline H$.
This emphasizes again the close connection between the two approaches,
although of course here the result here is not restricted to the NR regime.

The technical point just alluded to concerns the difference between the cutoff schemes
in the RFT and FVU approaches: the former using the smooth cutoff function $H(\vec k)$,
the latter a hard cutoff $\Lambda$.\footnote{%
In this regard, we note that the vertex functions in the FVU approach do not provide damping
factors  in the sums over spectator momenta, as can be seen for example by the
appearance of $v_\on^{-1}$ in $\widetilde C$.}
Thus, for example, $F_s$ from Eq.~(\ref{eq:F}) includes $H(\vec k)$, whereas
$D_L$ does not. Similarly there is a contribution to $\cK_{2,s}$ proportional to
$1-H(\vec k)$ that turns on near the cutoff. This implies that 
Eq.~(\ref{eq:DLb}) breaks down for $|\vec k| \sim m$, 
and thus that Eq.~(\ref{eq:MDQC3b}) is not strictly a valid rewriting of the FVU quantization
condition.
We view this, however, as a technical, and not a fundamental, issue.
The differences between cutoffs occur for spectator momenta such that
$E_{2,k}^{*2} \lesssim 0$.
In this regime, the three-particle threshold lies far away
(it opens up at $E_{2,k}^{*2}=4m^2$).
Thus the contributions to the sum over $\vec k$ with $E_{2,k}^{*2}\lesssim 0$ lead
only to exponentially-suppressed volume effects, and  we expect that varying
the cutoff in this regime can be compensated by changes to infinite-volume quantities,
namely the functions $C$ and $\Kdf$. This is exactly what happens in NREFT,
but in that case in a way that is simple to calculate.
Here the cutoff dependence of $C$ and $\Kdf$ will not be simple.
In light of these considerations, we will proceed using Eq.~(\ref{eq:MDQC3b})
as written.

Returning to the algebraic relation between the quantization conditions,
we begin from the form of the RFT quantization condition,
Eq.~(\ref{eq:HSQC2}), that looks most similar to the FVU result, Eq.~(\ref{eq:MDQC3b}).
As in the NREFT case, this similarity is superficial, because $\overline H^R$ in the
former equation and $\widetilde C/(2\omega L^3)$ here have very different properties.
However, if we assume that $\Kdfs$ is isotropic, $\Kdfs=|1\rangle \Kiso \langle 1 |$,
then, by essentially the same algebraic steps as earlier, we can show that the
two quantization conditions agree if
\begin{equation}
\widetilde C = \left[ 1 + (\cK_{2,s}^{-1} + I_s)\frac{ \Kdfs}{9L^3} \frac1{2\omega}\right]^{-1}
(\cK_{2s}^{-1}+ I_s) \frac{\Kdfs}{9} (\cK_{2,s}^{-1}+ I_s)\,.
\label{eq:CtoKdf}
\end{equation}
Here the relativistic generalization of the integral $I_s^\NR$ is
\begin{equation}
I_{s,pk} = \delta_{pk} \PV \int_{\vec a} \frac{H(\vec a)}{2 \omega_a ( [P-k - a]^2 - m^2 )}\,,
\label{eq:Is}
\end{equation}
where a specific choice for the regulator has been made.
Unlike in the NR regime, this integral does depend on $\vec k$.
Using the definition of $\widetilde C$, we can manipulate 
Eq.~(\ref{eq:CtoKdf}) into an equation for $C'$:
\begin{equation}
 C' = v_\on (\cK_{2s}^{-1}+ I_s) |1\rangle
 \frac1{1 + \Kiso \langle 1 | (2\omega)^{-1} (\cK_{2,s}^{-1} + I_s) | 1 \rangle/(9 L^3)}
\frac{\Kiso}{9} \langle 1 | (\cK_{2,s}^{-1}+ I_s) v_\on \,.
\label{eq:CptoKdf}
\end{equation}
This provides an explicit relation between the quantization conditions in the limit
of an isotropic $\Kdfs$. Note that 
\begin{equation}
\frac{\langle 1 | (2\omega)^{-1} (\cK_{2,s}^{-1} + I_s) | 1 \rangle}{L^3}
= \int_{\vec k} \frac{\cK_{2,s}(\vec k)^{-1} + I_s(\vec k) }{2\omega_k} + \cO( [\Lambda L]^{-1} )\,,
\end{equation}
so the right-hand side of Eq.~(\ref{eq:CptoKdf}) is volume independent  up to cutoff effects.

The most important lesson from Eq.~(\ref{eq:CptoKdf})
is that, if $\Kdfs$ is isotropic, then the function $C'$ cannot be.
The main reason for this is the presence of the vertex factors $v_\on$ in
Eq.~(\ref{eq:CptoKdf}). Recall that $v_\on$ is a diagonal matrix with entries
\begin{equation}
\lambda(E_{2,k}^{*2}) f(Q_k^2)\,,\quad Q_k^2 = E_{2,k}^{*2}-4 m^2\,.
\end{equation}
Thus it has a significant dependence on $\vec k$, with $f$ increasing as $|\vec k|$
becomes large and $Q_k^2$ becomes negative with a large magnitude.
Additional dependence on $\vec k$ enters through both $\cK_{2,s}$ and $I_s$.
This means that the vector
\begin{equation}
v_\on (\cK_{2,s}^{-1}+I_s) |1\rangle
\end{equation}
appearing on both ends of the expression for $C'$ is far from isotropic once
one leaves the NR regime.
We stress that this difference occurs not just near the cutoff, but rather over
the entire range of relativistic spectator momenta.

We do not think this conclusion is fundamentally surprising, since we do not
expect either $\Kdf$ or $C'$ to be close to isotropic once one leaves the NR regime.
Indeed, as shown in
Refs.~\cite{BBHRS,BHSnum}, one can systematically expand $\Kdf$ about threshold,
and only the leading two terms are isotropic. 
Presumably the same holds for $C'$.
Once in the relativistic domain one
needs the whole tower of higher-order terms, 
which are not isotropic (and also bring in higher-order dimers).

\subsubsection{Summary} The FVU method provides a direct and relatively simple
approach to obtain the three-particle quantization condition that is not restricted
to the NR regime. It follows steps that are,
roughly speaking, the relativized version of those used in the NREFT derivation.
The simplicity compared to the RFT approach is partly due to keeping only a single
dimer, but mainly because unitarity is used to determine which finite-volume momentum
sums have singular summands, rather than a diagrammatic analysis. 
Another advantage of the FVU approach is that it expected to work also in the
presence of subchannel resonances.

As noted above, we find the arguments for this approach very plausible, but
are not convinced that it constitutes a complete derivation.
Thus we think that it is important to show that it is equivalent to the result of
the RFT approach. We have taken the first steps in this direction above, showing
algebraic equivalence in the particular case of an isotropic $\Kdfs$.
Extending this analysis to more general forms for $\Kdf$ and to include higher-order
dimers is an important future challenge.

An important issue in this regard is whether the parametrization of $\cM_3$ given
in Ref.~\cite{MD0} is completely general. One way of checking this would be to
make a detailed comparison with the representation of $\cM_3$ 
in terms of $\Kdf$ that is a byproduct of the RFT approach 
(sketched in Sec.~\ref{sec:KtoM} above)~\cite{HS2}.
Since this representation follows from an all-orders diagrammatic approach,
it must be unitary, and indeed this can be shown explicitly~\cite{BHSS}.

\section{NUMERICAL\ IMPLEMENTATIONS}\label{sec:num}

In this section we give a brief summary of numerical results that have been calculated using each of the three formalisms presented above. In all cases, the results were determined by taking a model or ansatz for the infinite-volume interactions and then determining the corresponding finite-volume energy levels. We view these results as a proof of principle that, for all three formalisms, the mapping between finite-volume energies and infinite-volume scattering observables is feasible. A dedicated study in which only LQCD inputs are used to extract the three-body scattering amplitude has yet to be implemented, although Ref.~\cite{MD2} already made first steps in this direction.


Beginning with the RFT formalism described in Sec.~\ref{sec:QC3}, here we present two numerical results that are described in greater detail in Ref.~\cite{BHSnum}. In both calculations, the numerical evaluation was performed using the isotropic approximation, outlined in Sec.~\ref{sec:trunc}. We additionally considered systems for which the two-to-two scattering amplitude is well-described by the leading-order effective-range expansion, and thus depends only on the scattering length, $a$.

\begin{figure}
\begin{center}
\includegraphics[width=1\textwidth]{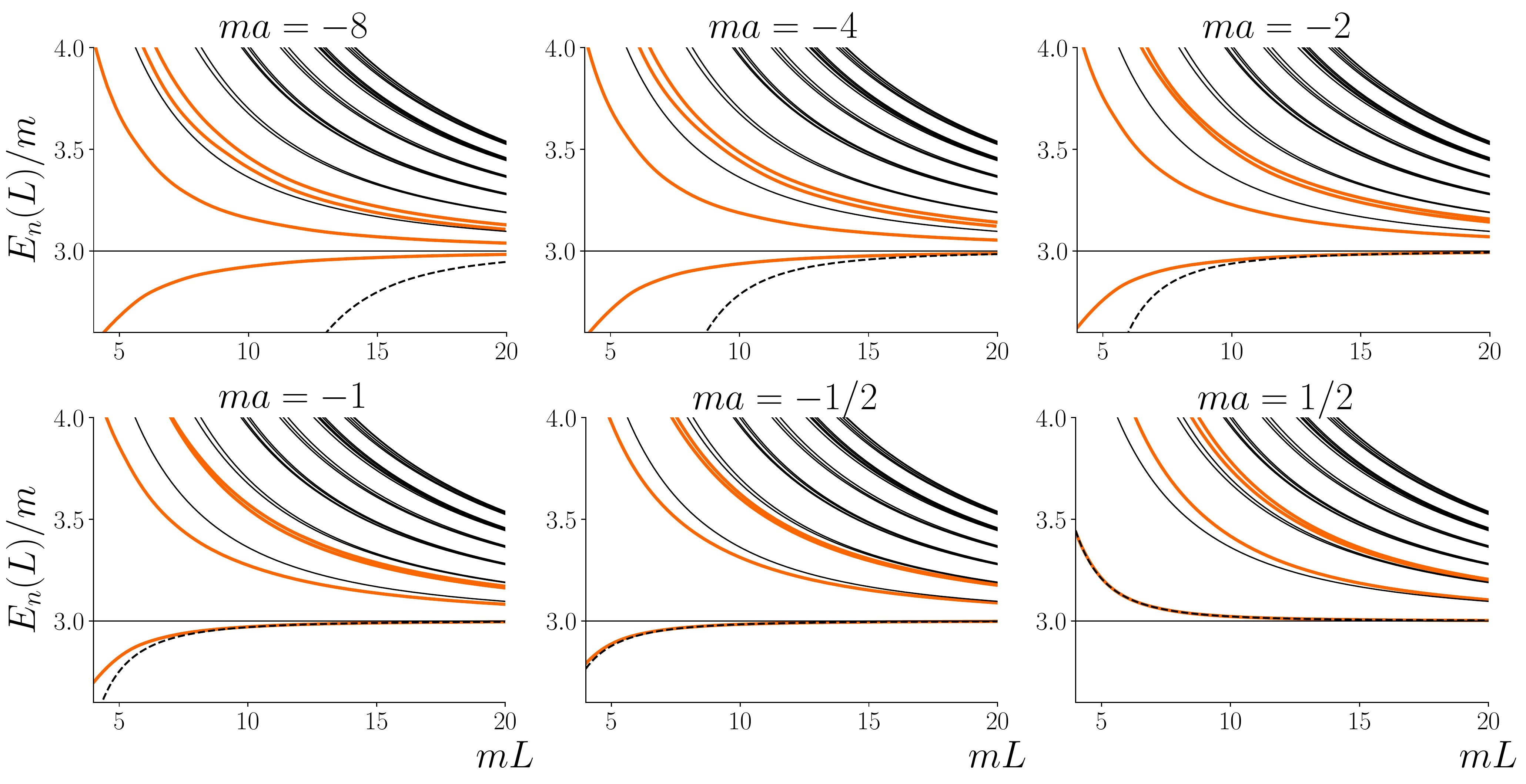}
\caption{Spectrum of three-particle states evaluated using the RFT formalism in the isotropic approximation, for various choices of the scattering length, $a$, and vanishing three-body interaction, $\mathcal K^{\text{iso}}_{\text{df},3}=0$. The thick orange curves show the interacting levels and the thin black lines show the corresponding energies of three non-interacting particles (i.e. the spectrum when $a=0$). In each plot, the dashed curve shows the $1/L$ expansion for the threshold state, Eq.~(\ref{eq:thresh_res}), which is
expected to work well when $a/L \ll 1$.  \label{fig:varya}}
\end{center}
\end{figure}

For our first application, we suppose that the local three-body interaction is negligible and set $\mathcal K_{\text{df},3}^{\text{iso}}=0$. Within this set-up, each choice of scattering length gives a prediction for the three-particle energies $E_n(L)$, as shown in {\bf Figure~\ref{fig:varya}}. These curves serve as a benchmark since, in future LQCD calculations, only by measuring deviations from the $\mathcal K_{\text{df},3} = 0$ predictions can one obtain information about three-body interactions.

\begin{figure}
\begin{center}
\includegraphics[width=\textwidth]{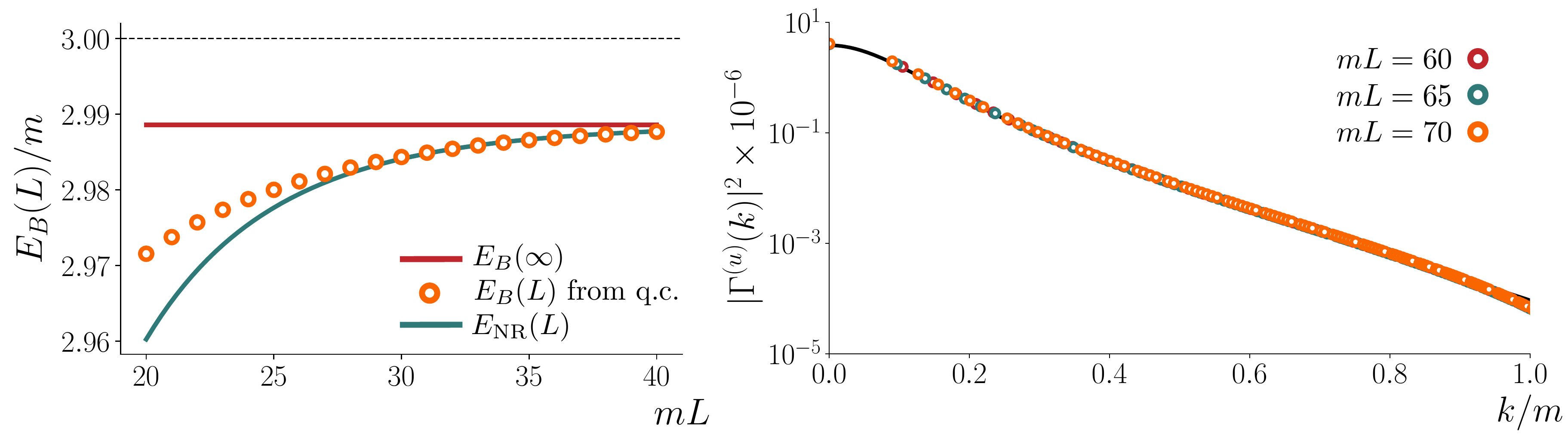}
\caption{Finite-volume bound-state energy (left) and the corresponding infinite-volume wave function (right). In the left panel we show the infinite-volume bound-state mass (horizontal red line) together with the leading-order non-relativistic prediction (green curve) and the result of numerically solving the RFT quantization condition (orange points). The asymptotic prediction agrees well with the numerical solution for $\kappa L>3$, but deviates for smaller volumes, as expected.
The right panel shows the result for the
residue of $\cM_3$ at the bound-state pole,
obtained by solving the integral equations relating $\Kdf$ to $\cM_3$.
We use a very large effective volume as a tool for numerically solving the infinite-volume equations and the consistency of the various data points indicates
 that we have reached the infinite-volume limit to high precision. 
 The solid black curve is an analytic prediction (not a fit to the data) given in Ref.~\cite{HS3bound}.   \label{fig:bound}}
\end{center}
\end{figure}

In the second numerical example from Ref.~\cite{BHSnum}, 
we work close to the unitary limit $1/(ma)=-10^{-4}$,
corresponding to a two-particle interaction that almost, but not quite, leads to a bound dimer.
In this limit, we find that, still working in the isotropic approximation, we can tune
$\mathcal K^{\text{iso}}_{\text{df},3}$ so that the infinite-volume system develops an
Efimov-like bound state~\cite{Efimov}.
In our example, the bound state energy is
$E_B \approx 2.99 m$, corresponding to a binding momentum of $\kappa \approx 0.1 m$. 
The quantization condition then determines the volume dependence of this state,
with results that are compared with the prediction of Ref.~\cite{MRR}, 
Eq.~(\ref{eq:MRRresult}), in the left panel of {\bf Figure~\ref{fig:bound}}.
We find good agreement for sufficiently large $L$, with
the single parameter determined to be $|A|^2\approx 0.95$.
This is in the expected range of values, i.e. close to unity.
 We expect, and find, that the curve deviates from the asymptotic form once $\kappa L  < 3$,
 because neglected $1/(\kappa L)$ corrections then become important. 
A key point, however, is that the quantization condition itself incorporates all terms suppressed by any power or any exponent of $\kappa L$. Its validity requires only
that $m L$ is sufficiently large. Thus one could, in this simple model, extract a value
for $\Kiso$ from the spectrum for $m L \approx 5$ (where typical LQCD calculations are done),
in a regime where the asymptotic formula, Eq.~(\ref{eq:MRRresult}), completely fails.

The right right panel of {\bf Figure~\ref{fig:bound}} shows a further test of the RFT formalism.
Here  we implemented our $\mathcal K_{\text{df},3} \Rightarrow \mathcal M_3$ relation
(described in Sec.~\ref{sec:KtoM}) to obtain the residue of $\cM_3$ at the bound-state pole.
 The resulting, numerically-determined residue is compared to the known analytic prediction 
 in the plot. This prediction is derived in Ref.~\cite{HS3bound} using NRQM, and, given the
 result for $|A|^2$ from the fit to the spectrum, is parameter-free.
 The good agreement  over seven orders of magnitude
 shows that this simple model of interactions captures the
 physics of the Efimov effect over a wide range of scales.

\begin{figure}
\begin{center}
\includegraphics[width=0.8\textwidth]{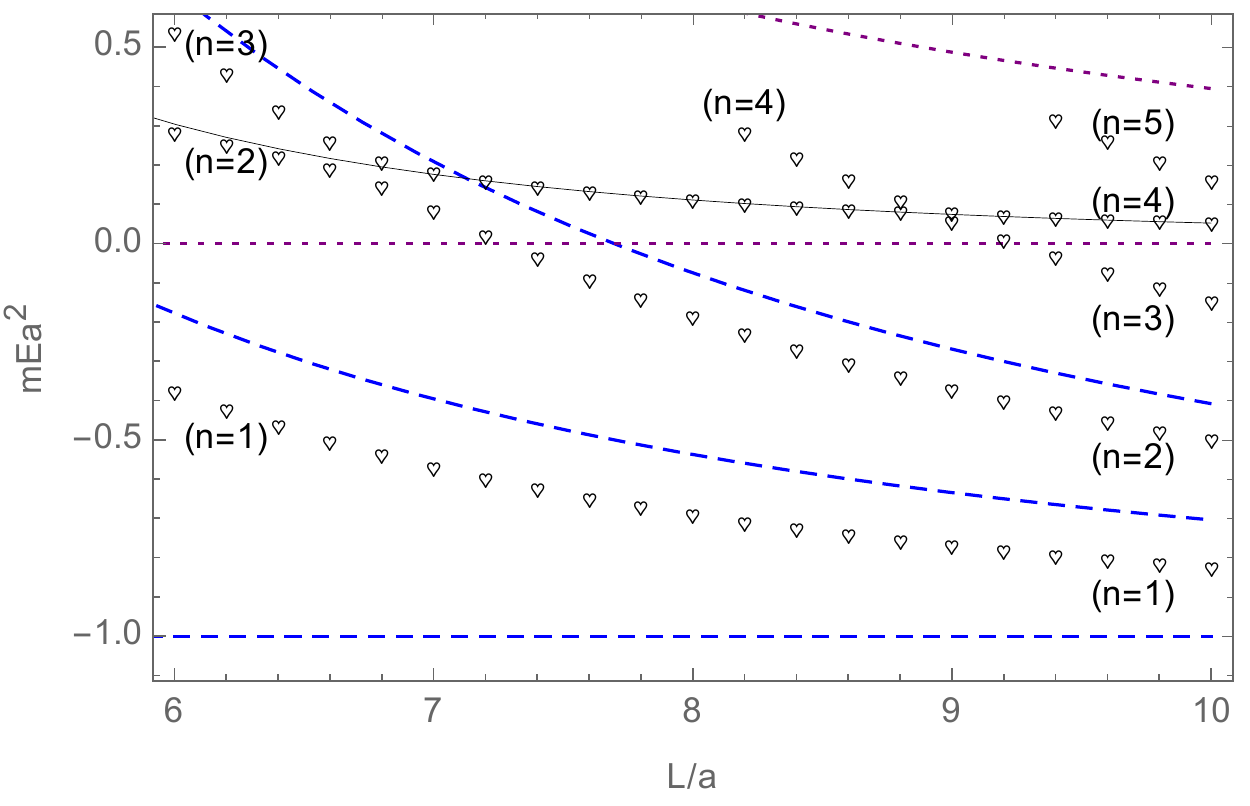}
\caption{Examle of three-particle energy levels obtained using the NREFT quantization condition
in Ref.~\cite{DHMPRW}. Here $E$ is the NR energy, called $E_{\text{NR}}$ elsewhere in this
review. The triangles show the results of solving the quantization condition for the parameters
described in the text. Only the states above the bound dimer-particle threshold at
$E_{\text{NR}}=-1/(ma^2)$ are shown (so the two bound trimer states lie below the bottom of the
plot and are not visible). 
The dashed blue lines show the energies of noninteracting dimer-particle states
with varying back-to-back momenta. The purple dotted curves show the free three-particle
states. The solid black line is the prediction of the threshold expansion,
 Eq.~(\ref{eq:thresh_res}), keeping terms only up to $\cO(1/L^5)$.
We observe that dimer-particle interactions push the energies up in finite volume,
as shown by the lowest two [($n=1$) and ($n=2$)] levels shown. The third level shown,
however, changes its nature as $L/a$ decreases: starting as a dimer-particle state,
converting to a three-particle threshold state at $L/a \sim 9$, and then converting back
to a different particle-dimer state at $L/a \sim 6.5$.
\label{fig:NREFT}}
\end{center}
\end{figure}

We next describe an example of the numerical results obtained using the NREFT approach,
as presented in Ref.~\cite{DHMPRW}. The setup is quite similar to that just described
for the RFT results, with the two-particle interactions described by an $s$-wave
scattering length, 
and a single isotropic three-body coupling constant. 
The total momentum is $\vec P=0$, and only the $A_1^+$ irrep is considered.
The scattering length is chosen
so that there is a bound dimer with NR energy, $E_{\text{NR}}= -1/(ma^2)$, and $H_0$ is
chosen so that there is a deeply bound trimer with $E_{\text{NR}} = -10/(ma^2)$.
It turns out that, with these parameters, there is, in infinite volume, a second trimer,
with $E_{\text{NR}} = -1.016/(ma^2)$, consisting of a loosely bound dimer and particle.
In finite volume, one then expects a plethora of states: the deeply bound trimer with
(asymptotically predicted) volume dependence, 
the second trimer with a different volume dependence,
a spectrum of states lying close to the
energies of a noninteracting dimer and particle,
and, finally, states that lie near the energies of three free particles.
An example of the resulting spectrum is shown in {\bf Figure~\ref{fig:NREFT}},
with only the non-bound-states shown. The expected states are seen, but with
a number of avoided level crossings making the interpretation nontrivial.
The overall conclusion is
that this rather complicated physical situation 
is successfully encoded into  the NREFT quantization condition.

\begin{figure}
\begin{center}
\includegraphics[width=0.8\textwidth]{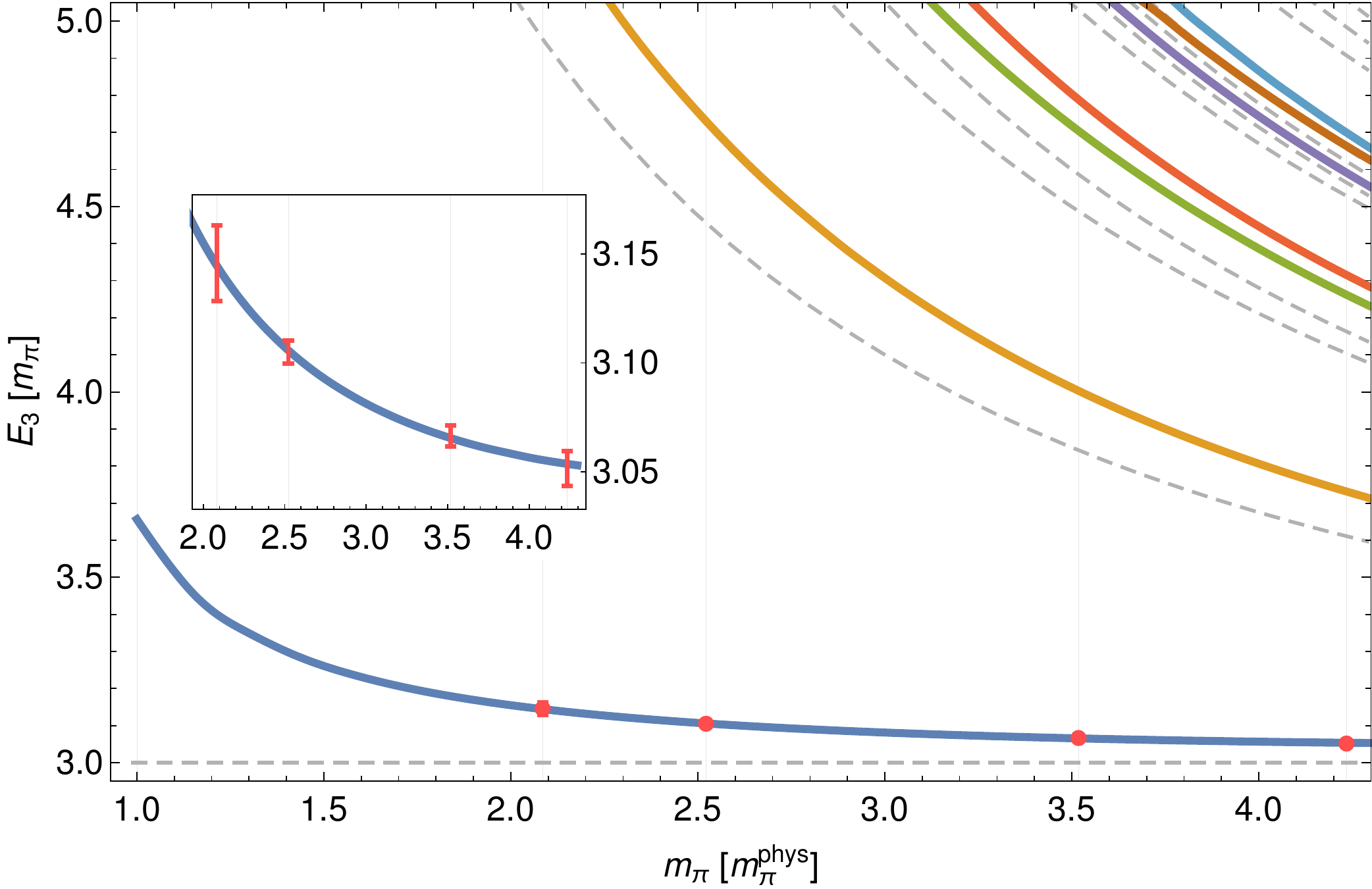}
\caption{Three-pion energy levels (solid lines, various colors)
presented in Ref.~\cite{MD2}.
The curves are based on two-to-two interactions determined from chiral perturbation theory combined with the inverse amplitude method. As in {\bf Figure~\ref{fig:varya}}, the three-body interaction term, $C$, is set to zero. 
In contrast to previous plots, in this case the interactions and the values of $m_\pi L$ are varying simultaneously, by changing the physical value of $m_\pi$ while keeping the box size fixed at $L=2.5\;$fm.
Only states in the $A_1^+$ irrep are shown, and 
the dashed lines show the noninteracting three-particle levels.
The inset zooms in on the fit that is used to constrain the three-particle interaction.
We caution that the lowest plotted values of $m_\pi$ correspond to $m_\pi L \approx 1.8$, for which the neglected, exponentially-suppressed terms may be significant.  \label{fig:MD}}
\end{center}
\end{figure}

Finally, we describe results from a recent numerical study of the FVU approach~\cite{MD2}. 
In this study the authors were motivated to make contact with a physical system by considering finite-volume $\pi^+ \pi^+$ as well as $\pi^+ \pi^+ \pi^+$ states.
They work in the isospin-symmetric limit, so these two sectors are not connected.
The $\pi^+ \pi^+ \to \pi^+ \pi^+$ scattering amplitude, needed as an input to the quantization
condition, 
was modeled by combining chiral perturbation theory amplitudes 
with the so-called inverse amplitude method. 
The resulting functional form also encodes pion-mass dependence, 
allowing $m_\pi$ to be varied from its physical value to a value around four times larger. 
In the energy range considered the approximation of keeping only $s$-wave dimers is 
expected to be very good.

This set-up enabled the authors to change $m_\pi L$ while holding the box size fixed at
 $L=2.5\;$fm, 
 with the resulting finite-volume energies varying due to the change in the effective box size (measured in units of $m_\pi$) as well as due to modifications in the interactions of the theory. 
 {\bf Figure~\ref{fig:MD}} shows the the resulting spectrum of $\pi^+ \pi^+ \pi^+$ states.
 For the lowest level (the threshold state) the energies can be compared to those
 previously determined by the NPLQCD collaboration in a 
 numerical LQCD calculation \cite{NPL1,NPL2}, allowing 
the three-body coupling $C$ to be constrained.
Using the (non-isotropic) form $C(\vec p, \vec q)= c \, \delta^3(\vec p - \vec q)$,
the result is consistent with zero, $c = (0.2 \pm 1.5) \times 10^{-10}$.

\begin{summary}[SUMMARY POINTS]
\begin{enumerate}
\item A method for determining predictions from lattice QCD (LQCD) for the properties
of resonances that have decay channels into three or more particles is urgently needed.
This will allow LQCD to address the nature of many of the higher-lying resonances,
in particular the recently observed $X$, $Y$ and $Z$ states.
\item To address this problem one requires a quantization condition that
relates the finite-volume spectrum of QCD, which can be obtained directly in
LQCD calculations, to the infinite-volume scattering amplitudes that encode resonance properties.
The two-particle quantization condition is known and widely used; 
this review focuses on progress
towards the three-particle quantization condition.
\item Three approaches have been used: one that is general and relativistic (RFT approach),
and also leads to a very complicated derivation; a second based in NREFT
that leads to a much simpler derivation; and a third that implements the constraints of unitarity in
the finite volume (FVU approach), which also leads to a simpler derivation
than the RFT approach but is nevertheless relativistic. 
The latter two approaches have only been formulated to date for $s$-wave dimer
interactions.
\item All three approaches lead to a two-step relation between the finite-volume spectrum and
infinite-volume scattering amplitudes, involving intermediate, cut-off dependent,
unphysical infinite-volume quantities.
\item The three approaches can be shown to be equivalent in certain regimes.
\item All three approaches have been successfully implemented numerically
in model calculations using the simplest approximations for interactions.
\end{enumerate}
\end{summary}

\begin{issues}[FUTURE ISSUES]
The development of the three-particle quantization condition has reached a pivotal stage.
The groundwork has been laid, but many technical issues must be addressed for the methodology to be applicable to most resonances of phenomenological interest.
Overall, we think that resolving these issues will be more straightforward than the
work that has been done so far, and thus we are optimistic about the future applicability
of the methodology.
We list here those issues that we consider most pressing. 
\begin{enumerate}
\item For all approaches, 
the formalism needs to be generalized to incorporate nonidentical particles and
particles with nonzero spin.
\item The NREFT and FVU approaches need to be extended to include dimers beyond
$s$-waves and moving frames, and to include the possibility of $2 \leftrightarrow 3$
transitions.
\item The NREFT approach needs to be ``relativized" in order to be applicable to
results from LQCD.
\item In the RFT approach, the second step of connecting
$\Kdf$ to $\cM_3$ must be implemented above threshold.
This work is further advanced in the NREFT and FVU approaches.
\item A technical issue in the RFT approach as presently formulated is the need to
use a relatively low cutoff (so that two-particle invariant masses are kept positive, i.e.~$E_{2,k}^{*2}\ge 0$). Extending the formalism to
allow a higher cutoff (concomitant with that used in the other approaches) should be
investigated. This is also related to an issue that we have not had space to discuss here,
namely whether the presence of the left-hand cut in the two-particle amplitude 
(which opens up at $E_{2,k}^{*2}=0$) can lead to difficulties for the formalisms.
\item Practical parametrizations of the three-particle interaction terms---$\Kdf$, $H(\Lambda)$
and $C$---must be developed that are based on phenomenological input and are flexible
enough to describe resonances encountered in the strong interactions.
\item The numerical implementations must be extended to include $2\leftrightarrow 3$
transitions. This is needed, for example, to study the Roper resonance.
\item The relation between the approaches must be studied when higher-spin dimers
and more complicated three-particle interactions are included.
\item Ultimately, the extension to four or more particles must be considered. This has
been worked out so far only for the energy of the
threshold state in NRQM~\cite{Detmold} and for the volume dependence of
an $N$-body NR bound state~\cite{KonigLee}.
\end{enumerate}
\end{issues}

\section*{DISCLOSURE STATEMENT}
The authors are not aware of any affiliations, memberships, funding, or financial holdings that
might be perceived as affecting the objectivity of this review. 

\section*{ACKNOWLEDGMENTS}
We are grateful to Michael D\"oring, Hans-Werner Hammer, Maxim Mai, and Akaki Rusetsky for helpful discussions and correspondence, and
to them, Jin-Yi Pang and J. Wu
for giving us permission to use plots from their work.
The work of SRS is supported in part by the United States Department of Energy grant 
No. DE-SC0011637.
 
%

%
%

%


\noindent

\end{document}